\documentclass[pre,twocolumn,floatfix, superscriptaddress,a4paper,nofootinbib]{revtex4-1}
\usepackage{amsmath,bm,graphicx}
\usepackage{amssymb}
\usepackage{amsfonts}
\usepackage{epstopdf}
\usepackage{mathrsfs}
\usepackage{xcolor}
\usepackage{latexsym}
\usepackage{hyperref}
\usepackage{subfigure}
\usepackage{bm}
\usepackage{natbib}
\usepackage{soul}
\usepackage[export]{adjustbox}

\newcommand{\comm}[1]{\textcolor{black}{#1}}
\newcommand{\pa}[1]{\textcolor{black}{#1}}

\newcommand{\Pe}{\operatorname{Pe}} 

\begin{document}

\title{Szilard engines
and information-based work extraction for active systems}

\author{Paolo Malgaretti}
\email[Corresponding Author: ]{p.malgaretti@fz-juelich.de }
\affiliation{Helmholtz Institute Erlangen-N\"urnberg for Renewable Energy (IEK-11), Forschungszentrum J\"ulich, Cauerstr. 1,
91058 Erlangen, Germany}

\author{Holger Stark}
\affiliation{Institut f\"ur Theoretische Physik, Technische Universit\"at Berlin, Hardenbergstr.~36, 10623 Berlin,\,Germany}

\begin{abstract}
The out of equilibrium nature of active systems can be exploited for the design of information-based engines.
We design two types of an active Szilard engine that use a Maxwell demon to extract work from an active bath composed of non-interacting Active Brownian Particles (ABPs). The two engines exploit either the quasi-static active pressure of ABPs or the long correlation time of their velocities.
\comm{For both engines the active bath allows to overcome the Landauer principle and to} extract larger work compared to conventional Szilard engines operating in quasi thermal equilibrium. For both of our engines,we identify the optimal regimes at which the work extracted and the efficiency are maximized. Finally, we discuss them in the context of synthetic and biological active systems.
\end{abstract}
\maketitle

Since Maxwell proposed his demon as 
\comm{an attempt to circumvent} the second law of thermodynamics~\cite{tait_book,Maxwell}, several attempts have been put forward  to rectify thermal fluctuations~\cite{Feynamn_book,Reimann_review}. 
An interesting twist to the idea of the demon has been provided by Szilard~\comm{\cite{Szilard1929,RexBook}}.
He proposed an engine that can extract work from a single thermal reservoir using information  obtained from the system when measuring some observable~\cite{Brillouin_book,Parrondo2015}.  
Recently, several works have addressed the realization of Szilard engines in experiments and theory spanning from the single electron~\cite{Koski2014} and boson systems~\cite{Reimann2018,Aydiner2021} up to the macromolecular~\cite{Ribezzi2019} and colloidal~\cite{Paneru2020} scale.

Up to now the ``fluctuations" in such Szilard engines have been regarded as thermal. 
\comm{This imposes constraints on the dynamics of the bath, 
which, among others, limit the extracted work by the Landauer principle.}  
However, in many situations the fluctuations are not of thermal origin. Rather, they are induced by the motion of active
agents,
such as swimming bacteria, a school of fish, or a swarm of drones
~\cite{Joanny_RMP,Ebbens_Review,Bechinger_RMP,Zoettl2016,Sagues_Review,Gompper_2020}.
\comm{This implies that 
Landauer's principle does not apply anymore raising
the question 
how well Szilard engines perform in contact with such non-thermal baths.}
\comm{The striking feature of 
active 
agents is 
their 
\comm{persistent}
motion, 
i.e., even within the overdamped regime their (active) velocity has a finite correlation time~\cite{Bechinger_RMP,Zoettl2016}, in contrast to the delta-correlated velocities of their passive counterparts. 
This is captured by the Active Brownian Particle (ABP) model~\cite{RomanczukSchimansky-Geier2012} characterized by a velocity 
correlation time $\tau_M$ and speed $v_0$ 
\cite{Zoettl2016}, which 
\comm{gives}
a persistence length $\lambda = v_0 \tau_M$. For 
\comm{$t \gg \tau_M$}
the motion of ABPs is diffusive with effective diffusion constant $D_{eff}=D+ v_0^2 \tau_M/d$ in $d$ dimensions 
\cite{Zoettl2016}, whereas for 
\comm{$t \lesssim \tau_M$}
their motion is persistent and it differs from their equilibrium counterparts. 
Despite its simplicity, such a simple model has been exploited to explain many experimentally observed phenomena spanning from wall accumulation\cite{Rotschild1963,ElgetiReview} to Motility Induced Phase Separation (MIPS)\cite{Cates13,Cates2015}. Indeed, recent works have addressed the role of a bath of active entities on the performance of heat-like \cite{Geisel2011,Fodor2019,Kroy2020,Malgaretti2021,Fodor2021,Gronchi2021,Speck2022,Santra2022}  and ratchet-like engines~\cite{Sokolov2010,DiLeonardo2010,Malgaretti2013JCP,
Malgaretti2014EPJ,Ripoll2014,Michelin2015,Malgaretti2017,Ryabov2017}.}

\comm{
In this letter we show that the persistent motion of active agents, modeled as  ABPs,
enables two generic realizations of an active Szilard engine.
First, in the traditional \textit{quasi-static} regime, 
the performance of the Szilard engine in contact with an active bath differs from its passive counterpart due to the active pressure~\cite{Malgaretti2021}. 
Second, 
we 
suggest a \textit{dynamic} Szilard engine that 
\comm{does not operate in the quasi-static regime but}
exploits the finite correlation time of the active velocity.}
In order to keep the model as general as possible, we rely on the minimal model
\comm{of an ABP to capture the effect of persistent motion.}
Indeed, such a simple (yet general) model allows for 
\comm{closed} 
expressions 
of the work extracted and its efficiency 
\comm{depending}
on the key parameters controlling the activity of the system. 
In particular, our active Szilard engines exhibit a remarkable improvement of the performance compared to their passive counterpart operating in thermal equilibrium, in line with recent experimental results~\cite{Zanin2021}. 
\begin{figure}
\centering
\includegraphics[scale=0.245,angle=0]{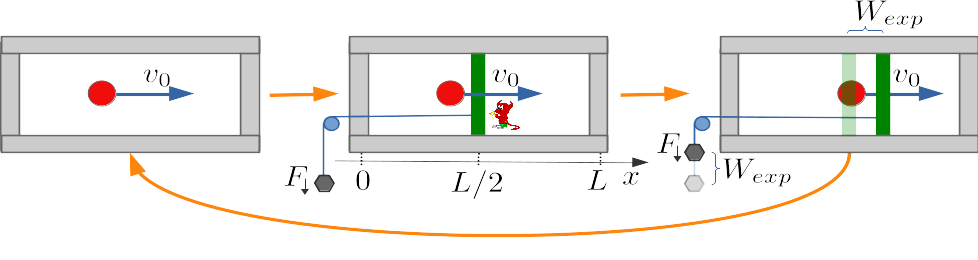}
\caption{\comm{Cartoon of the active Szilard engine. 
In the \textit{quasi-static} regime, the demon measures in which half of the box the active particle is
\comm{located}
and then 
\comm{places a  wall (in green) with an attached weight at $x = L/2$.}
After a full \textit{quasi-static} expansion performing work $W_{exp}$, the green wall hits the box
\comm{and is removed by the demon. This puts}
the system in its initial state. In the \textit{dynamic} regime, the 
demon measures the position of the particle with precision $\delta$ and 
the sign of the velocity. 
It puts a wall (in green) with precision $\delta$ ahead of the particle and let it push against 
the wall for a time $\tau$.
Then, the wall is removed and the system is back in its initial state.
}}
\label{fig:scheme0}
\end{figure}

\textit{Model}.
\comm{In the following, we focus on the motion of an overdamped ABP along the direction perpendicular to the wall and model its perpendicular velocity component}
as a random variable with \comm{finite} correlation time $\tau_M$ \comm{\cite{Zoettl2016}},
\comm{\begin{align}
\left\langle v(t)\right\rangle   =0\,, \quad 
\left\langle v(t)v(t')\right\rangle   = v_{0}^{2}e^{-\frac{|t-t'|}{\tau_{M}}} + 2D \delta(t-t')
\label{eq:v_corr_b}
\end{align}}
where 
\comm{$D$} 
is the \comm{(translational)} thermal diffusion coefficient. 
\comm{Equations (\ref{eq:v_corr_b}) have 
been shown to properly reproduce the statistical properties of ABPs (as modeled by Eqs.\ (S1) and (S2) in the Suppl. Mat.) as well 
as of Run-and-Tumble bacteria~\cite{Marconi2015}.}
The case of \comm{passive particles in}
thermal equilibrium is characterized by $v_0 = 0$.
\comm{In the active case and in the quasi-static regime ($t\gg \tau_M$) the model under study is equivalent to that of a particle moving in one dimension and hopping between two states, namely moving right or left~\cite{Malgaretti2021}.}
In the following, we analyze the performance of \comm{the two possible driving protocols of active  Szilard engines: the \textit{quasi-static} regime and the \textit{dynamic} regime.}

\textit{Quasi-static active Szilard engine}. 
\comm{First we focus on the case in which the engine moves on time scales larger than 
$\tau_M$ and hence the time evolution of the density of the ABPs can be regarded as quasi-static.}
The Maxwell demon in this type of Szilard engine operates as follows. 
\comm{A single ABP  is confined in a box with lateral area $A$ (see Fig.\ref{fig:scheme0}). 
The demon detects the position of the particle and inserts a wall in the middle of the box 
 ($x=L/2$). Due to the reduction of volume available to the particle, the pressure in the  
occupied chamber increases and the volume of the chamber expands till the movable wall hits the box (at $x= 0,L$). Then, the wall is removed and the system assumes its initial state again.}
  

\begin{figure}
\includegraphics[scale=0.33]{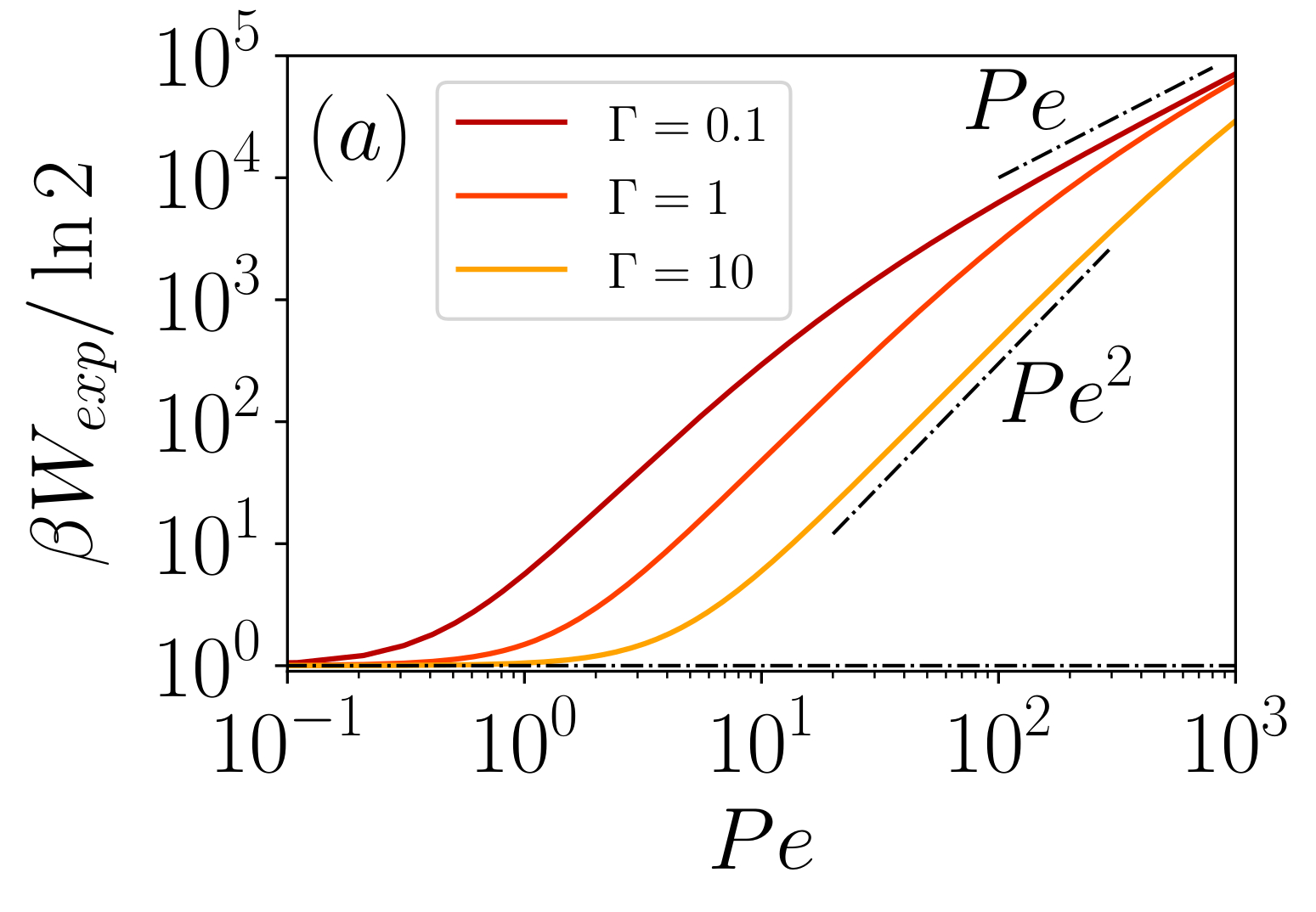}
\includegraphics[scale=0.33]{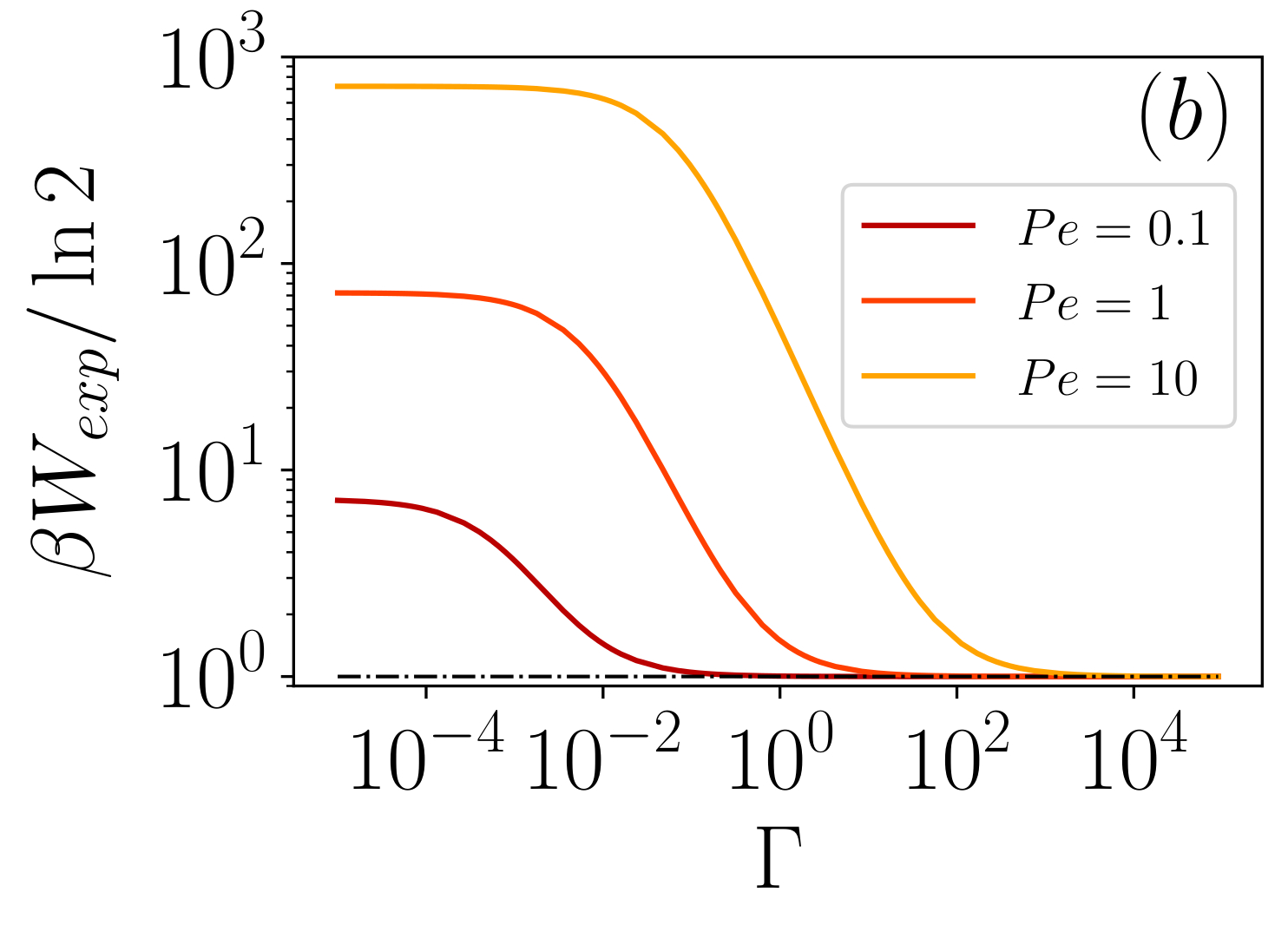}
\vspace{-17pt}
\caption{Work extracted during the expansion, $W_{exp}$, normalized by the equilibrium value of a Szilard engine, 
as function of Pe (a) or $\Gamma$ (b), with $L/R=100$.
\label{fig:W_sz}
}
\end{figure}

\comm{In the following, we consider just a single active particle\footnote{The same results holds also for the case of non-interacting active particles by simply multiplying 
the pressure by the number $N$ of active particles.} placed in the left half of the box as in Fig.\ \ref{fig:scheme0}.
The}
density is $\bar\rho=1/(Ax)$ 
and, under a mean-field approximation, the pressure is given by~\cite{Malgaretti2021}:

\begin{align}
\beta \Pi(x)=\frac{R^2}{A}\frac{\kappa^3}{\Pe^2+2\Gamma \kappa x}
\label{eq:Pi}
\end{align}
where
\begin{align}
\kappa=\frac{\sqrt{\Pe^2+2\Gamma}}{R},\,\,\Gamma=\frac{R^2}{\tau_M D},\,\,\Pe=\frac{v_0 R}{D} \, .
\end{align}
Here, $\kappa$ is the inverse of the length that characterizes the accumulation of 
ABPs at the walls~\cite{Malgaretti2021}
with $R$ the radius of the particle, $\Gamma$ is the dimensionless hopping rate, and $\Pe$
the P\'eclet number, respectively. Accordingly, the work extracted during the expansion becomes
%
\begin{align}
    \beta W_{exp}=A \int_{\frac{L}{2}}^{L} \beta \Pi(x) dx=\beta W_0\ln\left[\frac{1+\chi L}{1+\chi L/2}\right]\,,
   \label{eq:work-approx}
\end{align}
where $xA$ is the momentary volume available to the particle and we have identified the effective length $\chi^{-1}$ and strength $W_0$ of the work cycle,
\begin{align}
    \chi^{-1}=\frac{\Pe^2 R}{2\Gamma\sqrt{\Pe^2+2\Gamma}},&\quad 
\beta W_0 = 1 + \frac{\Pe^2}{2\Gamma}\,.
\end{align}
Figure\ \ref{fig:W_sz} shows the dependence of the work extracted from the system as function of $\Pe$ and $\Gamma$. 
As expected, in the limit of passive systems, i.e., for $\Pe\rightarrow 0$ or $\Gamma\rightarrow \infty$, $\beta W_{exp}$ approaches the work of the Szilard engine for a thermal bath $\beta W^{th}_{exp}=\ln 2$~\cite{Brillouin_book,Parrondo2015}.
At variance, for active systems ($\Pe>0$ and $\Gamma \ll \infty$), the work extracted in the case of an active bath exceeds that of the thermal bath. In particular, 
we can identify an intermediate regime, for $\sqrt{2\Gamma}<\Pe<2\Gamma L/R$ (see Ref.\cite{Malgaretti2021}), in which $\beta W_{exp}\propto \Pe^2$ whereas $\beta W_{exp}\propto \Pe$  in the asymptotic regime, $\Pe>2\Gamma L/R$.
\comm{Note}
that $W_{exp}\gg W_{exp}^{th}$ is not in contradiction with the Landauer principle~\cite{Landauer1961,Berut2012,Parrondo2015}
\comm{since the latter only applies to}
a thermal bath. 
In the case of active baths the work is done at the expense of the energy injected in the system to keep the bath in the active state.
\comm{The features we have described so far are generic for the active Szilard engine, namely the considerable excess work during 
expansion as compared to a thermal bath and its correct asymptotic behavior for $\Pe\rightarrow 0$ or $\Gamma\rightarrow \infty$.
However, the total extracted work during one cycle $W_{cyc}$ depends on the very realization of the active engine. 
In Suppl. Mat. Sec. B, we present an alternative protocol. It includes a compression step which results in a different dependence 
of $W_{cyc}$ on $\Pe$ and $\Gamma$ including local maxima.}

%
%

According to our model the energy input in the system over the cycling time $\tau$ is due to the measurement $\mathcal{M}$ and due to the energy  $\mathcal{W}\tau$ consumed, on average, by the ABP during the cycle period $\tau$ to keep itself in the active state. 
The input energy is partially dissipated and partially converted into the work, $W_{exp}$~\cite{VdBroek2007}.
Therefore we have $\mathcal{W}\tau +\mathcal{M} = W_{exp}+ W_{diss}$ with
\begin{align}
\label{eq.diss}
W_{diss} = \mathcal{P}\tau + \frac{\gamma\comm{_w} L^2}{\tau} +\mathcal{M} \, .
%
%
\end{align}
Note, since the measurement does not contribute to the work, we add it to the dissipated energy.
Furthermore, $\mathcal{P} \tau$ is the energy \textit{dissipated}, on average, by the active particle during the cycle period $\tau$ and $\gamma_w L^2/\tau$ is the (approximated) energy dissipated by the piston\footnote{For heat  engines that are in contact with thermal baths at different temperatures the heat exchange with the thermal baths should be added to $\langle W_{diss}\rangle$.}, where $\gamma_w$ is its effective friction coefficient.

Next, we define the efficiency~\cite{Schmiedl2007,Kajelstrup_book,Malgaretti2021} as 
\begin{equation}
\eta =\frac{W_{exp}}{\comm{W_{exp}}+W_{diss}} =
\left[{1+\dfrac{ W_{diss}}{ \comm{W_{exp}}  } }\right]^{-1}\,. 
\label{eq:eff}
\end{equation}
We remark that for the quasi-static processes 
assumed to derive Eq.~\eqref{eq:work-approx}, $ W_{exp}$ does not depend on $\tau$ and the dependence of $\eta$ on $\tau$ is determined solely by the dissipated energy $W_{diss}$. Interestingly, $W_{diss}$ diverges for both $\tau\rightarrow 0$  and $\tau\rightarrow \infty$ and it has a minimum at~\cite{Malgaretti2021} $\tau_{opt}=\sqrt{\gamma_w L^2/\mathcal{P}}$
which, via Eq.~\eqref{eq:eff}, implies a maximum for the efficiency. 
In order to be consistent with the quasi-static assumption, 
$\tau_{opt}$ should  exceed the typical relaxation time of the density profile 
that, for large $\text{Pe}$, scales as $\tau_{rlx}\simeq L/v_0$. 
Hence, $\tau_{rlx}<\tau_{opt}$ implies
\begin{align}
\label{eq:cond-max}
\gamma_w v_0^2>\mathcal{P} \, .
\end{align} 

For microswimmers it is well known that only a small part of the consumed mean power $\mathcal{W}$ can be converted
into work such as pushing a piston, while the rest is needed for the swim mechanism. Estimating this work power as 
$\gamma_{p} v_0^2$, where $\gamma_{p}$ is the friction coefficient of the ABP, the 
efficiency \comm{
\comm{$\alpha$}
of the ABP in a homogeneous and unbound fluid} 
becomes
\begin{align}
\alpha = \gamma_{p} v_0^2 / \mathcal{W} \ll 1\, .
\label{eq:def-alpha}
\end{align}
Reported values
for diffusiophoretic colloids are $\alpha\simeq 10^{-9}-10^{-5}$\ \cite{Benedikt2012,Shah2020} whereas for 
living organisms 
$\alpha\simeq 10^{-3}$ (\textit{Chlamydomonas}~\cite{Friederich2018}), and $\alpha\simeq 10^{-2}$ 
(\textit{Paramecium}~\cite{Katsu-Kimura2009} and
demembranated sperm  flagella~\cite{Daniel2015}).
Thus, with $\mathcal{W}\gtrsim\mathcal{P}$ and, using Eqs.~\eqref{eq:cond-max}-\eqref{eq:def-alpha}, we obtain
\begin{align}\label{eq:cond-max2}
\gamma_w \alpha > \gamma_{p} \,.
\end{align} 
Since both, $\gamma_p$ and $\gamma_w$, scale with their linear size, the constraint in Eq.~\eqref{eq:cond-max2}
together with $\alpha \ll 1$ requires a significant length scale separation between the ABP and the piston.
This is also due to the quasi-static process, we considered so far, where  pressure instantaneously relaxes to its stationary value. 
Therefore, now we formulate an engine where this constraint is released.

\textit{Dynamic active Szilard engine}
The intrinsic out-of-equilibrium  nature of active systems provides additional means to control their dynamics. For example, in the case of Active Brownian Particles (ABP) one can exploit the finite correlation time of their velocity to design a specific ``demon". Within such a scenario, the protocol of the demon works as follows. Every time $\tau$ the demon measures \comm{the position, with finite precision\footnote{\comm{The relation between the precision of the position measurement, $\delta$, and the measurement cost, $\mathcal{M}$, is not trivial due to the active nature of the ABP. We can estimate it from the thermal case where 
\comm{$\mathcal{M}\simeq -k_BT \ln(\delta/ 2L)$. The factor 2 comes from measuring the direction of motion}\cite{Brillouin_book}.}} 
$\delta$, and the direction of motion  of the ABP and it puts a wall ahead of the particle with precision $\delta$.} The wall is connected to a weight that opposes the motion of the particle with a force $F$ (see Fig.\ref{fig:scheme0}). 
Thus, \comm{after colliding with the wall, the ABP pushes against it 
\comm{and, hence,}  
acts against both the force $F$} 
and the drag force of the wall, $\gamma_w v_w$,
where $\gamma_w$ is the friction coefficient of the wall and $v_w$ its velocity.
Accordingly, at contact with the wall the ABP moves at a reduced velocity
\begin{align}\label{eq:barv}
v_w = \frac{\gamma_p v_0-F}{\gamma_p + \gamma_w} = v_0 \frac{1-F/\gamma_p v_0}{1 + \gamma_w/\gamma_p} \equiv v_0\bar{v}_w
\end{align}
with $\gamma_p$ the friction coefficient of the ABP.
Now, at time $t=0$ the velocity $v_w$ of the particle is measured while pushing against the wall.
Then, the work done against the conservative force $F$ in between two measurements  is 
\begin{equation}
W_{act}=F \int_{\comm{\tau_{\delta}}}^\tau v(t)\big |_{v_w}dt \, ,
\end{equation}
where $v(t)\big |_{v_w}$ means that at $t=0$ the velocity is $v_w$ 
\comm{and $\tau_\delta\simeq \delta/v_0$ is the time the particle takes to reach the wall placed with precision $\delta$.}
Ac\-cor\-dingly, in the case of large activity $v_w^2\gg D/\tau_M$ \comm{and relatively high measurement precision $\tau_\delta \ll \min(\tau_M,\tau)$,} 
the average work is linear in $\tau_M$ and becomes (see Suppl. Mat. Sec. B)
\begin{align}
\left \langle W_{act}\right\rangle \comm{\simeq} F v_w \tau_M \big (1-e^{-\bar\tau} \big) \, ,
\label{eq:W}
\end{align}
where we introduced the reduced measurement time $\bar{\tau}=\tau/\tau_M$. 
Introducing the rescaled work $\overline{W}_{act} = \langle W_{act} \rangle / (\gamma_p v_0^2 \tau_M)$, we realize that it depends on two dimensionless parameters, the reduced conservative force $\bar{F} = F / \gamma_pv_0$ and the ratio of drag coefficients
$\bar{\gamma}=\gamma_w/\gamma_p$.
\begin{figure}
\includegraphics[scale=0.33,valign=t]{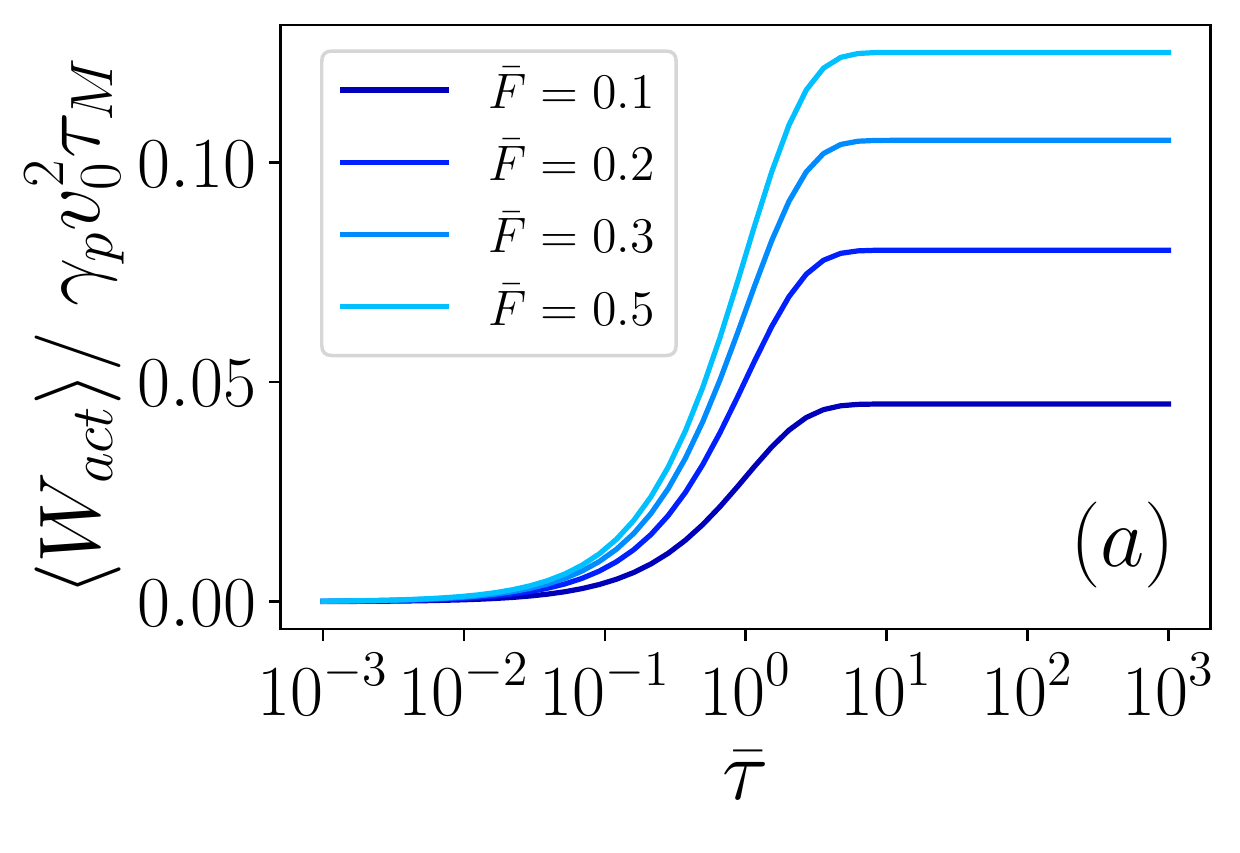}
\includegraphics[scale=0.33,valign=t]{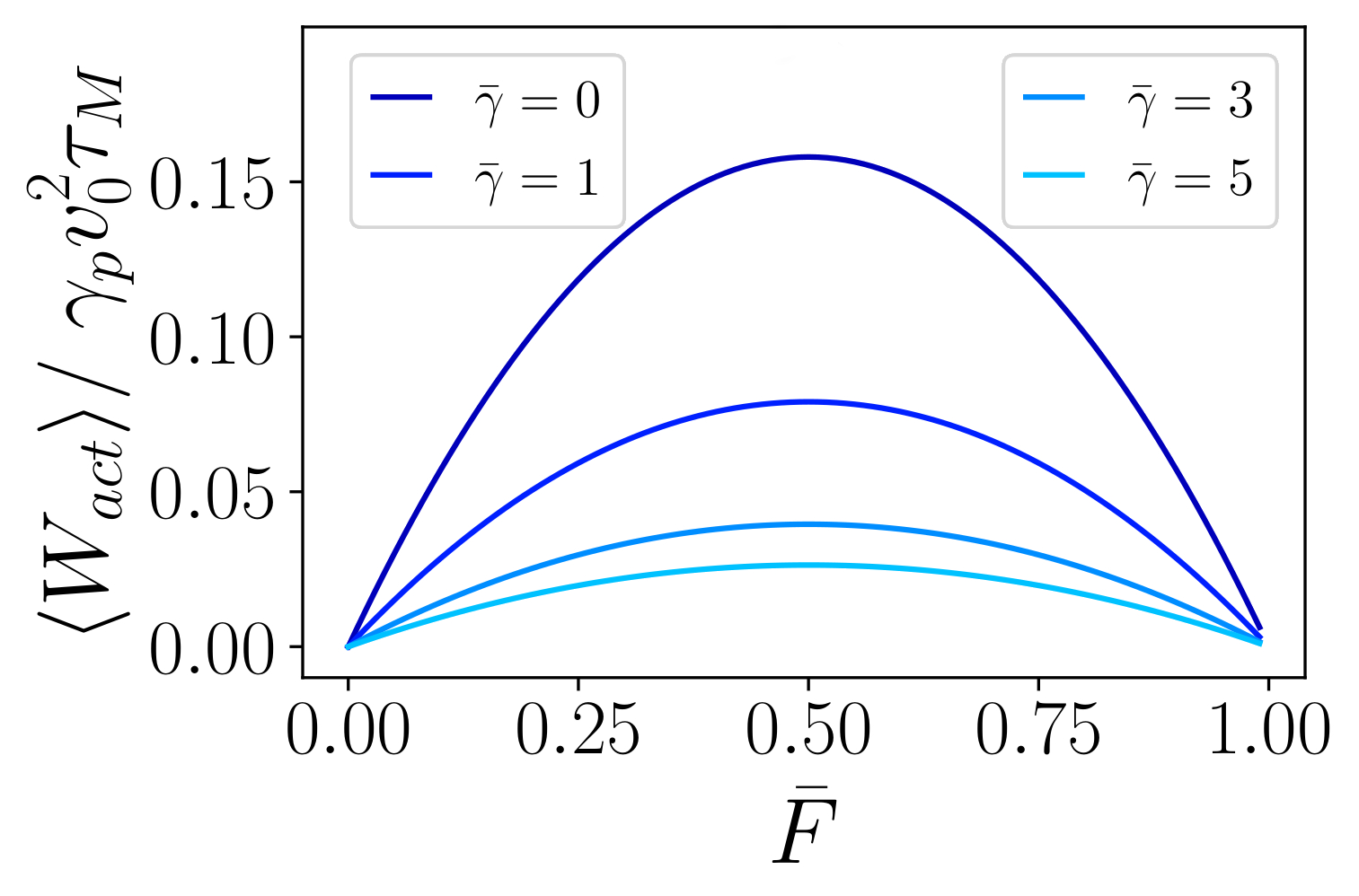}
\vspace{-11pt}
\caption{
Normalized work $\overline{W}_{act} = \langle W_{act}\rangle / (\gamma_p v_0^2 \tau_M)$ extracted during a measurement interval  of rescaled duration $\bar{\tau}$ as function of $\bar\tau$ with $\gamma_w/\gamma_p=1$ (a) or $\bar{F}$ with $\bar\tau=1$ (b).
\label{fig:Wact}}
\end{figure}
Figure~\ref{fig:Wact}(a) plots the dependence of the rescaled $\left\langle W_{act}\right\rangle$ on $\bar\tau$.  As expected, in the limit of thermal equilibrium, $\tau_M\rightarrow0$, $\left\langle W_{act}\right\rangle$ vanishes, whereas for $\tau\gg \tau_M$, $\langle \mathcal{W}_{act}\rangle$ attains its maximum value $F v_w \tau_M$, which in rescaled units becomes $\bar{F} \bar{v}_w$.
Due to Eq.\ (\ref{eq:barv}) for $\bar{v}_w$, $\left\langle W_{act}\right\rangle$ 
has a quadratic dependence on $\bar{F}$ [see Fig.\ \ref{fig:Wact}(b)] and it attains a maximum at $\bar{F}=1/2$, \emph{i.e.}, when the force exerted by the piston is half of the 
stall force of the ABP. The magnitude of $\beta \left\langle W_{act}\right\rangle$ 
can be estimated for an ABP with radius $R=\bar{R}\cdot 10^{-6}\mu\text{m}$, velocity $v_0=\bar{v}_0\cdot 10^{-6}\text{m/sec}$,
and moving in water with $\tau_M=1/D_r=8\pi\eta R^3/k_BT$ as $\beta \gamma_p v_0^2 \tau_M \simeq 25 \bar{R}^4 \bar{v}^2_0$. 
Accordingly, for $\bar{R}=2$ and $\bar{v}=2$ the work extracted at $\bar{F}=1/2$ is  $\beta \left\langle W_{act}\right\rangle 
\simeq 240$ whereas for biological swimmers it amounts to  $\beta \left\langle W_{act}\right\rangle \simeq 10^8$ for 
\textit{Clamydomonas} ($R=10\mu\text{m}$, $v=50\mu\text{m/sec}$, see Ref.\cite{Friederich2018}) and $\beta \left\langle 
W_{act}\right\rangle \simeq 10^{12}$ for \textit{Paramecium} ($R=25\mu\text{m}$, $v=10^3 \mu\text{m/sec}$, see 
Ref.\cite{Katsu-Kimura2009}).

Interestingly, in contrast to the quasi-static Szilard engine [see Eq.~\eqref{eq:work-approx}], in the current case the average work 
$\left\langle W_{act}\right\rangle$  retains an explicit dependence on $\tau$ that 
will be also visible in the efficiency.
In order to compute its mean value using the definition of 
Eq.~\eqref{eq:eff}, we need to introduce the total dissipated energy per time step
in full analogy to Eq.\ (\ref{eq.diss}),
\begin{align}
\left\langle  W_{diss}\right\rangle =\mathcal{P} \tau+\mathcal{M}+\left\langle  W^{pst}_{diss}\right\rangle  \, ,
\label{eq:W_irr2}
\end{align}
where 
\begin{align}\label{eq:W_diss_pst}
\left\langle  W^{pst}_{diss}\right\rangle = \frac{\gamma_w v_w^2 \tau_M}{2}  \overline{W}^{pst}_{diss}
\enspace \mathrm{with} \enspace
\overline{W}^{pst}_{diss} = 1- e^{-2\bar{\tau}}
%
%
\end{align}
is the energy dissipated by the piston (see Suppl. Mat. Sec.C).
In the limit of small measurement times,
$\bar\tau\ll 1$, it approaches $\left\langle  W^{pst}_{diss}\right\rangle = \gamma_w v_w^2 \tau$, as expected, while for long measurement times, $\bar\tau \gg 1$, it plateaus at $\left\langle  W^{pst}_{diss}\right\rangle = \gamma_w v_w^2 \tau_M/2$.
\begin{figure}[t]
\includegraphics[scale=0.43,valign=t]{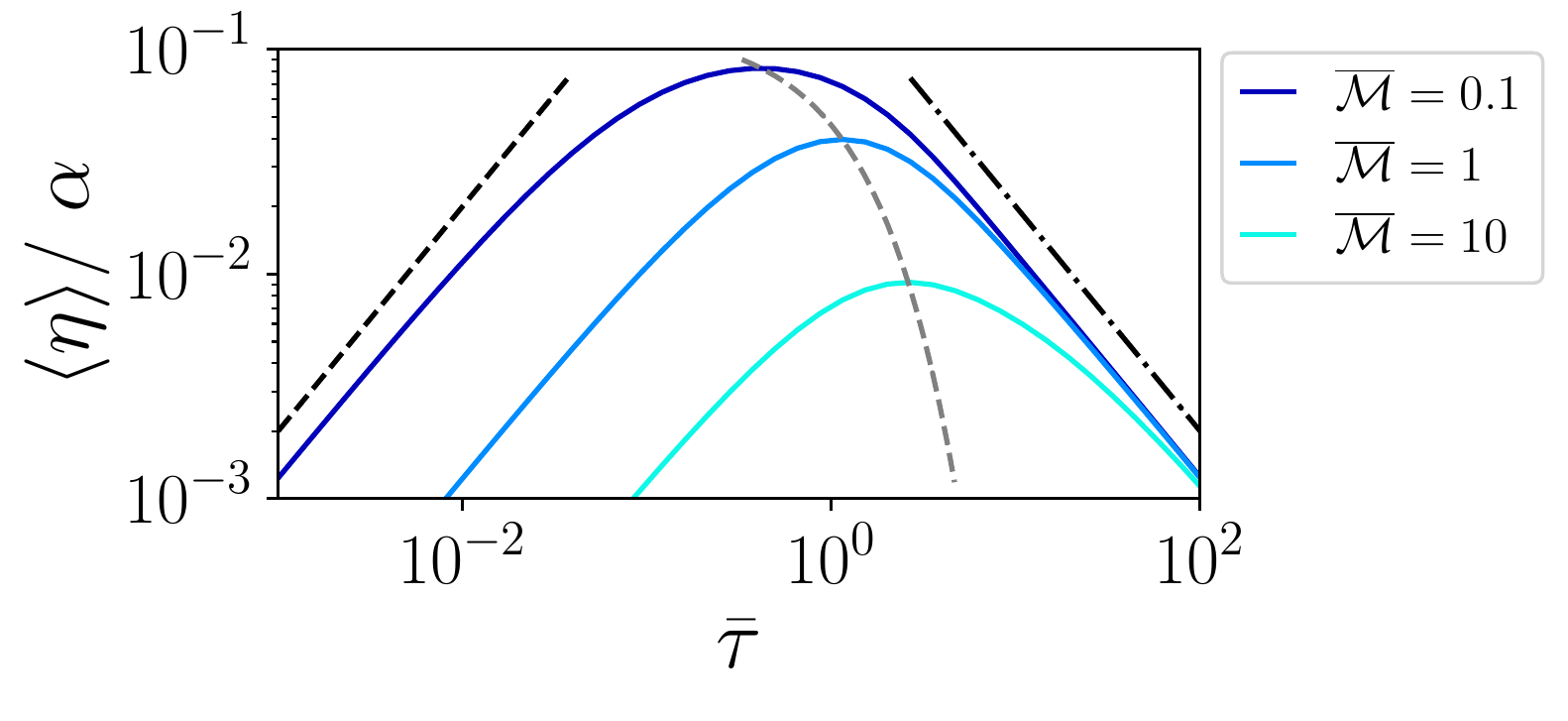}
\vspace{-11pt}
\caption{Efficiency $\left\langle \tilde\eta\right\rangle$ as function
of $\bar{\tau}$, for $\overline{\mathcal{M}}=0.1,1,10$ lighter colors standing for larger values of $\overline{\mathcal{M}}$. The black dot-dashed line is $\propto 1/\bar\tau$, the black dashed line is $\propto \bar\tau$, and the grey dashed line is the loci of the maxima for $\overline{\mathcal{M}} \in [0.01,100]$.
\vspace{-20pt}
 }
\label{fig:eff-1}
\end{figure}
Finally, the mean efficiency after some algebra (see Suppl. Mat. Sec.D), becomes 
\begin{align}
\left\langle \eta\right\rangle &\simeq \left[1+ \dfrac{1}{\alpha} \dfrac{\bar\tau+\overline{\mathcal{M}}+\frac{1}{2} \alpha \bar{\gamma} \bar{v}_w^2\overline{W}^{pst}_{\!diss}}{\overline{W}_{\!act}}\right]^{-1} 
\label{eq:eta}
\end{align}
where we approximated $\gamma_p v_0^2/\mathcal{P}\simeq \alpha$ and we have introduced the dimensionless  energy cost of measurement 
$\overline{\mathcal{M}}  =\frac{\mathcal{M}}{\mathcal{P} \tau_M}$.
\comm{Equation \eqref{eq:eta} can be further simplified since $\alpha \ll 1$ and 
$\bar{\gamma} \bar{v}_w^2\overline{W}^{pst}_{\!diss}\lesssim \bar\tau$, which follows from Eq.~\eqref{eq:W_diss_pst} and the definition of $\bar{\gamma}$ and $\bar{v}_{wall}$ (for more details, see Suppl. Mat. Sec. E). Hence, we obtain
}
\begin{align}
\langle \eta\rangle \simeq \alpha \dfrac{\overline{W}_{act}}{\bar\tau+\overline{\mathcal{M}}} \, .
\label{eq:tilde_eta}
\end{align}
Figure\ \ref{fig:eff-1} shows that the efficiency of the \textit{dynamic} Szilard engine, \comm{$\langle \eta\rangle$} is bound by that of the ABP, 
$\alpha$. Moreover,
\comm{$\langle \eta\rangle$} has a non-monotonous dependence on $\bar\tau$ \comm{independently whether the measurement cost is much smaller then ($\overline{\mathcal{M}}\ll 1$), similar to ($\overline{\mathcal{M}}\simeq 1$), or larger then ($\overline{\mathcal{M}}\gg 1$) the energy dissipated by the ABP to keep in the active state.} This is at variance to the monotonic increase of 
$\langle W_{act}\rangle$ in $\bar\tau$ [see Fig.\ \ref{fig:Wact}(a)] and is due to the linear increase of the power dissipated by 
the ABP with $\bar \tau$.
Interestingly,  the value, $\bar{\tau}_{opt}$, at which the efficiency is maximized grows with $\overline{\mathcal{M}}$ as $\bar{\tau}_{opt} \simeq \ln \overline{\mathcal{M}}$ (see the grey-dashed line in Fig.\ \ref{fig:eff-1} and Fig. S3), whereas the maximum value of \comm{$\langle \eta\rangle$} can reach up to $10\%$ of the efficiency of the ABP. \comm{Finally,  $\langle \eta\rangle $ depends on $\bar{F}$ solely through $\langle W_{act}\rangle$ and hence it retains the non-monotonous dependence on $\bar{F}$ [see Fig.\ref{fig:Wact}(b)].}

\comm{In this article we have discussed the general features of Szilard engines operating in contact with active baths. Importantly, the active bath differs from a thermal bath by the persistent motion of its constituents. The resulting enhanced active pressure and the exponential time correlations of the velocity enable two designs of an active Szilard engine that exploit different regimes.}


\comm{
In the \textit{quasi-static} regime the slow 
expansion of the active Szilard engine on times much larger than the typical relaxation time of the ABP's density is driven by the active pressure. Its magnitude depends on both the P\'eclet number as well as the box size~\cite{Malgaretti2021}. The }
work extracted from \comm{the}
active bath during expansion outnumbers the equilibrium counterpart.
Thus, the Landauer limit~\cite{Landauer1961,Berut2012,Parrondo2015} can be overcome at the expense of the additional energy pumped into the system by the active bath.
Since the active bath undergoes a \textit{quasi-static} work cycle, \comm{$W_{exp}$} is independent of the cycling time $\tau$. 
\comm{The efficiency depends on $\tau$ solely through} the energy dissipated by the ABP and the piston such that it becomes maximum at a finite value $\tau_{opt}$. 
For longer cycling times, the dissipation is essentially due to the energy spent to keep the system active and the efficiency scales $\propto 1/\tau$.

\comm{In the \textit{dynamic} regime the active Szilard engine  probes the velocity correlation time, $\tau_M$, of the ABP.
This implementation of the active Szilard engine has no \textit{practical} counterpart for thermal baths. 
In fact, for an ABP with radius $R=1\mu\text{m}$ suspended in water, we have $\tau_M\simeq D_r^{-1}=8\pi\eta R^3/k_BT\simeq 6\text{sec}$. In contrast, for a passive colloid of the same size and also suspended in water the velocity correlation time is $\tau_{th}\simeq m/\gamma_p\simeq 10^{-7}\text{sec}$, where $m$ is the mass of the colloidal particle. 
Accordingly, while an ABP can perform work 
on a relatively slow time scale of seconds, for a passive particle the demon  would need to monitor velocity correlations 
on a very fast time scale of $10^{-7}\sec$.}
Interestingly, the work extracted, $\langle {W}_{act}\rangle$,  becomes maximum when $F$ is half of the stall force $F_{opt}=\gamma_p v_0/2$. 
Since the extracted work scales as $\beta \langle {W}_{act} \rangle \propto R^4 v_0^2$, it varies in a large range from $1$ for micron-sized colloids up to $10^{12}$ for a \emph{Paramecium}.
The dynamic Szilard engine works with an efficiency $\langle \eta \rangle$ bounded from above by the efficiency $\alpha$ of an ABP.
Interestingly, $\langle \eta \rangle$ shows a non-monotonous dependence on both the dimensionless measuring time 
$\bar\tau=\tau/\tau_M$ with 
\comm{a maximum at $\bar\tau_{opt} \approx 1$,}
as well as on the conservative force with the maximum attained at $F_{opt}=\gamma_p v_0/2$.
For biological swimmers $\alpha \in [10^{-4},10^{-2}]$, hence according to Fig.\ \ref{fig:eff-1} the efficiency can be as high as 
$\langle \eta \rangle \simeq 10^{-3}$, which is quite larger than typical efficiencies of other micrometric engines with values smaller than $\sim 10^{-8}$ (see the diverse systems discussed in Ref.\ \cite{Maggi2015}).

\comm{Our results highlight how the persistent motion of active bath particles and the 
resulting features of enhanced  active pressure and long-time velocity or orientational correlations,  determine the dynamics of information-based engines. These features} 
can be exploited to design novel micro- and nano-engines that outperform those relying on equilibrium baths.
Using the method of optical video microscopy and real-time image analysis (see, \emph{e.g.}, \cite{Baeuerle2020}),
we envisage the possibility to mimic the Maxwell demon and to ultimately realize an active Szilard engine.

\section*{Aknowledgments}
P.M. acknoledges Wendong Wang, Gaurav Gardi, and Vimal Kishore for useful discussions.

\bibliography{daemon_biblio}

\begin{thebibliography}{54}%
\makeatletter
\providecommand \@ifxundefined [1]{%
 \@ifx{#1\undefined}
}%
\providecommand \@ifnum [1]{%
 \ifnum #1\expandafter \@firstoftwo
 \else \expandafter \@secondoftwo
 \fi
}%
\providecommand \@ifx [1]{%
 \ifx #1\expandafter \@firstoftwo
 \else \expandafter \@secondoftwo
 \fi
}%
\providecommand \natexlab [1]{#1}%
\providecommand \enquote  [1]{``#1''}%
\providecommand \bibnamefont  [1]{#1}%
\providecommand \bibfnamefont [1]{#1}%
\providecommand \citenamefont [1]{#1}%
\providecommand \href@noop [0]{\@secondoftwo}%
\providecommand \href [0]{\begingroup \@sanitize@url \@href}%
\providecommand \@href[1]{\@@startlink{#1}\@@href}%
\providecommand \@@href[1]{\endgroup#1\@@endlink}%
\providecommand \@sanitize@url [0]{\catcode `\\12\catcode `\$12\catcode
  `\&12\catcode `\#12\catcode `\^12\catcode `\_12\catcode `\%12\relax}%
\providecommand \@@startlink[1]{}%
\providecommand \@@endlink[0]{}%
\providecommand \url  [0]{\begingroup\@sanitize@url \@url }%
\providecommand \@url [1]{\endgroup\@href {#1}{\urlprefix }}%
\providecommand \urlprefix  [0]{URL }%
\providecommand \Eprint [0]{\href }%
\providecommand \doibase [0]{http://dx.doi.org/}%
\providecommand \selectlanguage [0]{\@gobble}%
\providecommand \bibinfo  [0]{\@secondoftwo}%
\providecommand \bibfield  [0]{\@secondoftwo}%
\providecommand \translation [1]{[#1]}%
\providecommand \BibitemOpen [0]{}%
\providecommand \bibitemStop [0]{}%
\providecommand \bibitemNoStop [0]{.\EOS\space}%
\providecommand \EOS [0]{\spacefactor3000\relax}%
\providecommand \BibitemShut  [1]{\csname bibitem#1\endcsname}%
\let\auto@bib@innerbib\@empty
\bibitem [{tai(1867)}]{tait_book}%
  \BibitemOpen
  \href@noop {} {\emph {\bibinfo {title} {Life and Scientific Work of Peter
  Guthrie Tait}}}\ (\bibinfo  {publisher} {Cambridge University Press},\
  \bibinfo {year} {1867})\ p.\ \bibinfo {pages} {213–215}\BibitemShut
  {NoStop}%
\bibitem [{\citenamefont {Maxwell}(1871)}]{Maxwell}%
  \BibitemOpen
  \bibfield  {author} {\bibinfo {author} {\bibfnamefont {J.}~\bibnamefont
  {Maxwell}},\ }\href@noop {} {\emph {\bibinfo {title} {Theory of Heat}}}\
  (\bibinfo  {publisher} {Longmans, Green and Co.},\ \bibinfo {address}
  {London},\ \bibinfo {year} {1871})\ Chap.~\bibinfo {chapter} {12}\BibitemShut
  {NoStop}%
\bibitem [{\citenamefont {Feynman}(1963)}]{Feynamn_book}%
  \BibitemOpen
  \bibfield  {author} {\bibinfo {author} {\bibfnamefont {R.}~\bibnamefont
  {Feynman}},\ }\href@noop {} {\emph {\bibinfo {title} {The Feynman Lectures on
  Physics, Vol. 1}}}\ (\bibinfo {year} {1963})\BibitemShut {NoStop}%
\bibitem [{\citenamefont {Reimann}(2002)}]{Reimann_review}%
  \BibitemOpen
  \bibfield  {author} {\bibinfo {author} {\bibfnamefont {P.}~\bibnamefont
  {Reimann}},\ }\href@noop {} {\bibfield  {journal} {\bibinfo  {journal} {Phys.
  Rep.}\ }\textbf {\bibinfo {volume} {361}},\ \bibinfo {pages} {57} (\bibinfo
  {year} {2002})}\BibitemShut {NoStop}%
\bibitem [{\citenamefont {Szilard}(1929)}]{Szilard1929}%
  \BibitemOpen
  \bibfield  {author} {\bibinfo {author} {\bibfnamefont {L.}~\bibnamefont
  {Szilard}},\ }\href@noop {} {\  (\bibinfo {year} {1929})}\BibitemShut
  {NoStop}%
\bibitem [{\citenamefont {Leff}\ and\ \citenamefont {Rex}(2002)}]{RexBook}%
  \BibitemOpen
  \bibfield  {author} {\bibinfo {author} {\bibfnamefont {H.}~\bibnamefont
  {Leff}}\ and\ \bibinfo {author} {\bibfnamefont {A.~F.}\ \bibnamefont {Rex}},\
  }\href@noop {} {\emph {\bibinfo {title} {Maxwell’s Demon 2}}}\ (\bibinfo
  {publisher} {Tyalor \& Francis},\ \bibinfo {address} {Bristol},\ \bibinfo
  {year} {2002})\BibitemShut {NoStop}%
\bibitem [{\citenamefont {Brillouin}(1960)}]{Brillouin_book}%
  \BibitemOpen
  \bibfield  {author} {\bibinfo {author} {\bibfnamefont {L.}~\bibnamefont
  {Brillouin}},\ }\href@noop {} {\emph {\bibinfo {title} {Science and
  Information Theory, 2nd Ed.}}}\ (\bibinfo  {publisher} {Dover Publications},\
  \bibinfo {address} {New York},\ \bibinfo {year} {1960})\BibitemShut {NoStop}%
\bibitem [{\citenamefont {Parrondo}\ \emph {et~al.}(2015)\citenamefont
  {Parrondo}, \citenamefont {Horowitz},\ and\ \citenamefont
  {Sagawa}}]{Parrondo2015}%
  \BibitemOpen
  \bibfield  {author} {\bibinfo {author} {\bibfnamefont {J.~M.~R.}\
  \bibnamefont {Parrondo}}, \bibinfo {author} {\bibfnamefont {J.~M.}\
  \bibnamefont {Horowitz}}, \ and\ \bibinfo {author} {\bibfnamefont
  {T.}~\bibnamefont {Sagawa}},\ }\href {\doibase 10.1038/nphys3230} {\bibfield
  {journal} {\bibinfo  {journal} {Nature Physics}\ }\textbf {\bibinfo {volume}
  {11}},\ \bibinfo {pages} {131} (\bibinfo {year} {2015})}\BibitemShut
  {NoStop}%
\bibitem [{\citenamefont {Koski}\ \emph {et~al.}(2014)\citenamefont {Koski},
  \citenamefont {Maisi}, \citenamefont {Pekola},\ and\ \citenamefont
  {Averin}}]{Koski2014}%
  \BibitemOpen
  \bibfield  {author} {\bibinfo {author} {\bibfnamefont {J.~V.}\ \bibnamefont
  {Koski}}, \bibinfo {author} {\bibfnamefont {V.~F.}\ \bibnamefont {Maisi}},
  \bibinfo {author} {\bibfnamefont {J.~P.}\ \bibnamefont {Pekola}}, \ and\
  \bibinfo {author} {\bibfnamefont {D.~V.}\ \bibnamefont {Averin}},\ }\href
  {\doibase 10.1073/pnas.1406966111} {\bibfield  {journal} {\bibinfo  {journal}
  {Proceedings of the National Academy of Sciences}\ }\textbf {\bibinfo
  {volume} {111}},\ \bibinfo {pages} {13786} (\bibinfo {year}
  {2014})}\BibitemShut {NoStop}%
\bibitem [{\citenamefont {Bengtsson}\ \emph {et~al.}(2018)\citenamefont
  {Bengtsson}, \citenamefont {Tengstrand}, \citenamefont {Wacker},
  \citenamefont {Samuelsson}, \citenamefont {Ueda}, \citenamefont {Linke},\
  and\ \citenamefont {Reimann}}]{Reimann2018}%
  \BibitemOpen
  \bibfield  {author} {\bibinfo {author} {\bibfnamefont {J.}~\bibnamefont
  {Bengtsson}}, \bibinfo {author} {\bibfnamefont {M.~N.}\ \bibnamefont
  {Tengstrand}}, \bibinfo {author} {\bibfnamefont {A.}~\bibnamefont {Wacker}},
  \bibinfo {author} {\bibfnamefont {P.}~\bibnamefont {Samuelsson}}, \bibinfo
  {author} {\bibfnamefont {M.}~\bibnamefont {Ueda}}, \bibinfo {author}
  {\bibfnamefont {H.}~\bibnamefont {Linke}}, \ and\ \bibinfo {author}
  {\bibfnamefont {S.~M.}\ \bibnamefont {Reimann}},\ }\href {\doibase
  10.1103/PhysRevLett.120.100601} {\bibfield  {journal} {\bibinfo  {journal}
  {Phys. Rev. Lett.}\ }\textbf {\bibinfo {volume} {120}},\ \bibinfo {pages}
  {100601} (\bibinfo {year} {2018})}\BibitemShut {NoStop}%
\bibitem [{\citenamefont {Aydiner}(2021)}]{Aydiner2021}%
  \BibitemOpen
  \bibfield  {author} {\bibinfo {author} {\bibfnamefont {E.}~\bibnamefont
  {Aydiner}},\ }\href {\doibase 10.1038/s41598-020-80639-w} {\bibfield
  {journal} {\bibinfo  {journal} {Scientific Reports}\ }\textbf {\bibinfo
  {volume} {11}},\ \bibinfo {pages} {1576} (\bibinfo {year}
  {2021})}\BibitemShut {NoStop}%
\bibitem [{\citenamefont {Ribezzi-Crivellari}\ and\ \citenamefont
  {Ritort}(2019)}]{Ribezzi2019}%
  \BibitemOpen
  \bibfield  {author} {\bibinfo {author} {\bibfnamefont {M.}~\bibnamefont
  {Ribezzi-Crivellari}}\ and\ \bibinfo {author} {\bibfnamefont
  {F.}~\bibnamefont {Ritort}},\ }\href {\doibase 10.1038/s41567-019-0481-0}
  {\bibfield  {journal} {\bibinfo  {journal} {Nature Physics}\ }\textbf
  {\bibinfo {volume} {15}},\ \bibinfo {pages} {660} (\bibinfo {year}
  {2019})}\BibitemShut {NoStop}%
\bibitem [{\citenamefont {Paneru}\ and\ \citenamefont
  {Pak}(2020)}]{Paneru2020}%
  \BibitemOpen
  \bibfield  {author} {\bibinfo {author} {\bibfnamefont {G.}~\bibnamefont
  {Paneru}}\ and\ \bibinfo {author} {\bibfnamefont {H.~K.}\ \bibnamefont
  {Pak}},\ }\href {\doibase 10.1080/23746149.2020.1823880} {\bibfield
  {journal} {\bibinfo  {journal} {Advances in Physics: X}\ }\textbf {\bibinfo
  {volume} {5}},\ \bibinfo {pages} {1823880} (\bibinfo {year}
  {2020})}\BibitemShut {NoStop}%
\bibitem [{\citenamefont {Marchetti}\ \emph {et~al.}(2013)\citenamefont
  {Marchetti}, \citenamefont {Joanny}, \citenamefont {Ramaswamy}, \citenamefont
  {Liverpool}, \citenamefont {Prost}, \citenamefont {Rao},\ and\ \citenamefont
  {Simha}}]{Joanny_RMP}%
  \BibitemOpen
  \bibfield  {author} {\bibinfo {author} {\bibfnamefont {M.~C.}\ \bibnamefont
  {Marchetti}}, \bibinfo {author} {\bibfnamefont {J.~F.}\ \bibnamefont
  {Joanny}}, \bibinfo {author} {\bibfnamefont {S.}~\bibnamefont {Ramaswamy}},
  \bibinfo {author} {\bibfnamefont {T.~B.}\ \bibnamefont {Liverpool}}, \bibinfo
  {author} {\bibfnamefont {J.}~\bibnamefont {Prost}}, \bibinfo {author}
  {\bibfnamefont {M.}~\bibnamefont {Rao}}, \ and\ \bibinfo {author}
  {\bibfnamefont {R.~A.}\ \bibnamefont {Simha}},\ }\href {\doibase
  10.1103/RevModPhys.85.1143} {\bibfield  {journal} {\bibinfo  {journal} {Rev.
  Mod. Phys.}\ }\textbf {\bibinfo {volume} {85}},\ \bibinfo {pages} {1143}
  (\bibinfo {year} {2013})}\BibitemShut {NoStop}%
\bibitem [{\citenamefont {Ebbens}\ and\ \citenamefont
  {Howse}(2010)}]{Ebbens_Review}%
  \BibitemOpen
  \bibfield  {author} {\bibinfo {author} {\bibfnamefont {S.~J.}\ \bibnamefont
  {Ebbens}}\ and\ \bibinfo {author} {\bibfnamefont {J.~R.}\ \bibnamefont
  {Howse}},\ }\href {\doibase 10.1039/B918598D} {\bibfield  {journal} {\bibinfo
   {journal} {Soft Matter}\ }\textbf {\bibinfo {volume} {6}},\ \bibinfo {pages}
  {726} (\bibinfo {year} {2010})}\BibitemShut {NoStop}%
\bibitem [{\citenamefont {Bechinger}\ \emph {et~al.}(2016)\citenamefont
  {Bechinger}, \citenamefont {Di~Leonardo}, \citenamefont {L\"owen},
  \citenamefont {Reichhardt}, \citenamefont {Volpe},\ and\ \citenamefont
  {Volpe}}]{Bechinger_RMP}%
  \BibitemOpen
  \bibfield  {author} {\bibinfo {author} {\bibfnamefont {C.}~\bibnamefont
  {Bechinger}}, \bibinfo {author} {\bibfnamefont {R.}~\bibnamefont
  {Di~Leonardo}}, \bibinfo {author} {\bibfnamefont {H.}~\bibnamefont
  {L\"owen}}, \bibinfo {author} {\bibfnamefont {C.}~\bibnamefont {Reichhardt}},
  \bibinfo {author} {\bibfnamefont {G.}~\bibnamefont {Volpe}}, \ and\ \bibinfo
  {author} {\bibfnamefont {G.}~\bibnamefont {Volpe}},\ }\href {\doibase
  10.1103/RevModPhys.88.045006} {\bibfield  {journal} {\bibinfo  {journal}
  {Rev. Mod. Phys.}\ }\textbf {\bibinfo {volume} {88}},\ \bibinfo {pages}
  {045006} (\bibinfo {year} {2016})}\BibitemShut {NoStop}%
\bibitem [{\citenamefont {Zöttl}\ and\ \citenamefont
  {Stark}(2016)}]{Zoettl2016}%
  \BibitemOpen
  \bibfield  {author} {\bibinfo {author} {\bibfnamefont {A.}~\bibnamefont
  {Zöttl}}\ and\ \bibinfo {author} {\bibfnamefont {H.}~\bibnamefont {Stark}},\
  }\href {\doibase 10.1088/0953-8984/28/25/253001} {\bibfield  {journal}
  {\bibinfo  {journal} {Journal of Physics: Condensed Matter}\ }\textbf
  {\bibinfo {volume} {28}},\ \bibinfo {pages} {253001} (\bibinfo {year}
  {2016})}\BibitemShut {NoStop}%
\bibitem [{\citenamefont {Doostmohammadi}\ \emph {et~al.}(2018)\citenamefont
  {Doostmohammadi}, \citenamefont {Ign{\'e}s-Mullol}, \citenamefont {Yeomans},\
  and\ \citenamefont {Sagu{\'e}s}}]{Sagues_Review}%
  \BibitemOpen
  \bibfield  {author} {\bibinfo {author} {\bibfnamefont {A.}~\bibnamefont
  {Doostmohammadi}}, \bibinfo {author} {\bibfnamefont {J.}~\bibnamefont
  {Ign{\'e}s-Mullol}}, \bibinfo {author} {\bibfnamefont {J.~M.}\ \bibnamefont
  {Yeomans}}, \ and\ \bibinfo {author} {\bibfnamefont {F.}~\bibnamefont
  {Sagu{\'e}s}},\ }\href {\doibase 10.1038/s41467-018-05666-8} {\bibfield
  {journal} {\bibinfo  {journal} {Nature Communications}\ }\textbf {\bibinfo
  {volume} {9}},\ \bibinfo {pages} {3246} (\bibinfo {year} {2018})}\BibitemShut
  {NoStop}%
\bibitem [{\citenamefont {Gompper}\ \emph {et~al.}(2020)\citenamefont
  {Gompper}, \citenamefont {Winkler}, \citenamefont {Speck}, \citenamefont
  {Solon}, \citenamefont {Nardini}, \citenamefont {Peruani}, \citenamefont
  {Löwen}, \citenamefont {Golestanian}, \citenamefont {Kaupp}, \citenamefont
  {Alvarez}, \citenamefont {Ki{\o}rboe}, \citenamefont {Lauga}, \citenamefont
  {Poon}, \citenamefont {DeSimone}, \citenamefont {Mui{\~{n}}os-Landin},
  \citenamefont {Fischer}, \citenamefont {Söker}, \citenamefont {Cichos},
  \citenamefont {Kapral}, \citenamefont {Gaspard}, \citenamefont {Ripoll},
  \citenamefont {Sagues}, \citenamefont {Doostmohammadi}, \citenamefont
  {Yeomans}, \citenamefont {Aranson}, \citenamefont {Bechinger}, \citenamefont
  {Stark}, \citenamefont {Hemelrijk}, \citenamefont {Nedelec}, \citenamefont
  {Sarkar}, \citenamefont {Aryaksama}, \citenamefont {Lacroix}, \citenamefont
  {Duclos}, \citenamefont {Yashunsky}, \citenamefont {Silberzan}, \citenamefont
  {Arroyo},\ and\ \citenamefont {Kale}}]{Gompper_2020}%
  \BibitemOpen
  \bibfield  {author} {\bibinfo {author} {\bibfnamefont {G.}~\bibnamefont
  {Gompper}}, \bibinfo {author} {\bibfnamefont {R.~G.}\ \bibnamefont
  {Winkler}}, \bibinfo {author} {\bibfnamefont {T.}~\bibnamefont {Speck}},
  \bibinfo {author} {\bibfnamefont {A.}~\bibnamefont {Solon}}, \bibinfo
  {author} {\bibfnamefont {C.}~\bibnamefont {Nardini}}, \bibinfo {author}
  {\bibfnamefont {F.}~\bibnamefont {Peruani}}, \bibinfo {author} {\bibfnamefont
  {H.}~\bibnamefont {Löwen}}, \bibinfo {author} {\bibfnamefont
  {R.}~\bibnamefont {Golestanian}}, \bibinfo {author} {\bibfnamefont {U.~B.}\
  \bibnamefont {Kaupp}}, \bibinfo {author} {\bibfnamefont {L.}~\bibnamefont
  {Alvarez}}, \bibinfo {author} {\bibfnamefont {T.}~\bibnamefont {Ki{\o}rboe}},
  \bibinfo {author} {\bibfnamefont {E.}~\bibnamefont {Lauga}}, \bibinfo
  {author} {\bibfnamefont {W.~C.~K.}\ \bibnamefont {Poon}}, \bibinfo {author}
  {\bibfnamefont {A.}~\bibnamefont {DeSimone}}, \bibinfo {author}
  {\bibfnamefont {S.}~\bibnamefont {Mui{\~{n}}os-Landin}}, \bibinfo {author}
  {\bibfnamefont {A.}~\bibnamefont {Fischer}}, \bibinfo {author} {\bibfnamefont
  {N.~A.}\ \bibnamefont {Söker}}, \bibinfo {author} {\bibfnamefont
  {F.}~\bibnamefont {Cichos}}, \bibinfo {author} {\bibfnamefont
  {R.}~\bibnamefont {Kapral}}, \bibinfo {author} {\bibfnamefont
  {P.}~\bibnamefont {Gaspard}}, \bibinfo {author} {\bibfnamefont
  {M.}~\bibnamefont {Ripoll}}, \bibinfo {author} {\bibfnamefont
  {F.}~\bibnamefont {Sagues}}, \bibinfo {author} {\bibfnamefont
  {A.}~\bibnamefont {Doostmohammadi}}, \bibinfo {author} {\bibfnamefont
  {J.~M.}\ \bibnamefont {Yeomans}}, \bibinfo {author} {\bibfnamefont {I.~S.}\
  \bibnamefont {Aranson}}, \bibinfo {author} {\bibfnamefont {C.}~\bibnamefont
  {Bechinger}}, \bibinfo {author} {\bibfnamefont {H.}~\bibnamefont {Stark}},
  \bibinfo {author} {\bibfnamefont {C.~K.}\ \bibnamefont {Hemelrijk}}, \bibinfo
  {author} {\bibfnamefont {F.~J.}\ \bibnamefont {Nedelec}}, \bibinfo {author}
  {\bibfnamefont {T.}~\bibnamefont {Sarkar}}, \bibinfo {author} {\bibfnamefont
  {T.}~\bibnamefont {Aryaksama}}, \bibinfo {author} {\bibfnamefont
  {M.}~\bibnamefont {Lacroix}}, \bibinfo {author} {\bibfnamefont
  {G.}~\bibnamefont {Duclos}}, \bibinfo {author} {\bibfnamefont
  {V.}~\bibnamefont {Yashunsky}}, \bibinfo {author} {\bibfnamefont
  {P.}~\bibnamefont {Silberzan}}, \bibinfo {author} {\bibfnamefont
  {M.}~\bibnamefont {Arroyo}}, \ and\ \bibinfo {author} {\bibfnamefont
  {S.}~\bibnamefont {Kale}},\ }\href {\doibase 10.1088/1361-648x/ab6348}
  {\bibfield  {journal} {\bibinfo  {journal} {Journal of Physics: Condensed
  Matter}\ }\textbf {\bibinfo {volume} {32}},\ \bibinfo {pages} {193001}
  (\bibinfo {year} {2020})}\BibitemShut {NoStop}%
\bibitem [{\citenamefont {Romanczuk}\ \emph {et~al.}(2012)\citenamefont
  {Romanczuk}, \citenamefont {B{\"a}r}, \citenamefont {Ebeling}, \citenamefont
  {Lindner},\ and\ \citenamefont
  {Schimansky-Geier}}]{RomanczukSchimansky-Geier2012}%
  \BibitemOpen
  \bibfield  {author} {\bibinfo {author} {\bibfnamefont {P.}~\bibnamefont
  {Romanczuk}}, \bibinfo {author} {\bibfnamefont {M.}~\bibnamefont {B{\"a}r}},
  \bibinfo {author} {\bibfnamefont {W.}~\bibnamefont {Ebeling}}, \bibinfo
  {author} {\bibfnamefont {B.}~\bibnamefont {Lindner}}, \ and\ \bibinfo
  {author} {\bibfnamefont {L.}~\bibnamefont {Schimansky-Geier}},\ }\href
  {\doibase 10.1140/epjst/e2012-01529-y} {\bibfield  {journal} {\bibinfo
  {journal} {Eur. Phys. J. Spec. Top.}\ }\textbf {\bibinfo {volume} {202}},\
  \bibinfo {pages} {1} (\bibinfo {year} {2012})}\BibitemShut {NoStop}%
\bibitem [{\citenamefont {Rotschild}(1963)}]{Rotschild1963}%
  \BibitemOpen
  \bibfield  {author} {\bibinfo {author} {\bibfnamefont {L.}~\bibnamefont
  {Rotschild}},\ }\href@noop {} {\bibfield  {journal} {\bibinfo  {journal}
  {Nature}\ ,\ \bibinfo {pages} {198}} (\bibinfo {year} {1963})}\BibitemShut
  {NoStop}%
\bibitem [{\citenamefont {Elgeti}\ \emph {et~al.}(2015)\citenamefont {Elgeti},
  \citenamefont {Winkler},\ and\ \citenamefont {Gompper}}]{ElgetiReview}%
  \BibitemOpen
  \bibfield  {author} {\bibinfo {author} {\bibfnamefont {J.}~\bibnamefont
  {Elgeti}}, \bibinfo {author} {\bibfnamefont {R.~G.}\ \bibnamefont {Winkler}},
  \ and\ \bibinfo {author} {\bibfnamefont {G.}~\bibnamefont {Gompper}},\ }\href
  {\doibase 10.1088/0034-4885/78/5/056601} {\bibfield  {journal} {\bibinfo
  {journal} {Rep. Prog. Phys.}\ }\textbf {\bibinfo {volume} {78}},\ \bibinfo
  {pages} {056601} (\bibinfo {year} {2015})}\BibitemShut {NoStop}%
\bibitem [{\citenamefont {Cates}\ and\ \citenamefont
  {Tailleur}(2013)}]{Cates13}%
  \BibitemOpen
  \bibfield  {author} {\bibinfo {author} {\bibfnamefont {M.~E.}\ \bibnamefont
  {Cates}}\ and\ \bibinfo {author} {\bibfnamefont {J.}~\bibnamefont
  {Tailleur}},\ }\href@noop {} {\bibfield  {journal} {\bibinfo  {journal} {EPL
  (Europhys. Lett.)}\ }\textbf {\bibinfo {volume} {101}},\ \bibinfo {pages}
  {20010} (\bibinfo {year} {2013})}\BibitemShut {NoStop}%
\bibitem [{\citenamefont {Cates}\ and\ \citenamefont
  {Tailleur}(2015)}]{Cates2015}%
  \BibitemOpen
  \bibfield  {author} {\bibinfo {author} {\bibfnamefont {M.~E.}\ \bibnamefont
  {Cates}}\ and\ \bibinfo {author} {\bibfnamefont {J.}~\bibnamefont
  {Tailleur}},\ }\href {\doibase 10.1146/annurev-conmatphys-031214-014710}
  {\bibfield  {journal} {\bibinfo  {journal} {Annu. Rev. Condens. Matter
  Phys.}\ }\textbf {\bibinfo {volume} {6}},\ \bibinfo {pages} {219} (\bibinfo
  {year} {2015})}\BibitemShut {NoStop}%
\bibitem [{\citenamefont {Zheng}\ \emph {et~al.}(2011)\citenamefont {Zheng},
  \citenamefont {Ellis}, \citenamefont {Kottos}, \citenamefont {Fleischmann},
  \citenamefont {Geisel},\ and\ \citenamefont {Prosen}}]{Geisel2011}%
  \BibitemOpen
  \bibfield  {author} {\bibinfo {author} {\bibfnamefont {M.~C.}\ \bibnamefont
  {Zheng}}, \bibinfo {author} {\bibfnamefont {F.~M.}\ \bibnamefont {Ellis}},
  \bibinfo {author} {\bibfnamefont {T.}~\bibnamefont {Kottos}}, \bibinfo
  {author} {\bibfnamefont {R.}~\bibnamefont {Fleischmann}}, \bibinfo {author}
  {\bibfnamefont {T.}~\bibnamefont {Geisel}}, \ and\ \bibinfo {author}
  {\bibfnamefont {T.~c.~v.}\ \bibnamefont {Prosen}},\ }\href {\doibase
  10.1103/PhysRevE.84.021119} {\bibfield  {journal} {\bibinfo  {journal} {Phys.
  Rev. E}\ }\textbf {\bibinfo {volume} {84}},\ \bibinfo {pages} {021119}
  (\bibinfo {year} {2011})}\BibitemShut {NoStop}%
\bibitem [{\citenamefont {Pietzonka}\ \emph {et~al.}(2019)\citenamefont
  {Pietzonka}, \citenamefont {Fodor}, \citenamefont {Lohrmann}, \citenamefont
  {Cates},\ and\ \citenamefont {Seifert}}]{Fodor2019}%
  \BibitemOpen
  \bibfield  {author} {\bibinfo {author} {\bibfnamefont {P.}~\bibnamefont
  {Pietzonka}}, \bibinfo {author} {\bibfnamefont {{\'E}.}~\bibnamefont
  {Fodor}}, \bibinfo {author} {\bibfnamefont {C.}~\bibnamefont {Lohrmann}},
  \bibinfo {author} {\bibfnamefont {M.~E.}\ \bibnamefont {Cates}}, \ and\
  \bibinfo {author} {\bibfnamefont {U.}~\bibnamefont {Seifert}},\ }\href
  {\doibase 10.1103/PhysRevX.9.041032} {\bibfield  {journal} {\bibinfo
  {journal} {Phys. Rev. X}\ }\textbf {\bibinfo {volume} {9}},\ \bibinfo {pages}
  {041032} (\bibinfo {year} {2019})}\BibitemShut {NoStop}%
\bibitem [{\citenamefont {Holubec}\ \emph {et~al.}(2020)\citenamefont
  {Holubec}, \citenamefont {Steffenoni}, \citenamefont {Falasco},\ and\
  \citenamefont {Kroy}}]{Kroy2020}%
  \BibitemOpen
  \bibfield  {author} {\bibinfo {author} {\bibfnamefont {V.}~\bibnamefont
  {Holubec}}, \bibinfo {author} {\bibfnamefont {S.}~\bibnamefont {Steffenoni}},
  \bibinfo {author} {\bibfnamefont {G.}~\bibnamefont {Falasco}}, \ and\
  \bibinfo {author} {\bibfnamefont {K.}~\bibnamefont {Kroy}},\ }\href {\doibase
  10.1103/PhysRevResearch.2.043262} {\bibfield  {journal} {\bibinfo  {journal}
  {Phys. Rev. Res.}\ }\textbf {\bibinfo {volume} {2}},\ \bibinfo {pages}
  {043262} (\bibinfo {year} {2020})}\BibitemShut {NoStop}%
\bibitem [{\citenamefont {Malgaretti}\ \emph {et~al.}(2021)\citenamefont
  {Malgaretti}, \citenamefont {Nowakowski},\ and\ \citenamefont
  {Stark}}]{Malgaretti2021}%
  \BibitemOpen
  \bibfield  {author} {\bibinfo {author} {\bibfnamefont {P.}~\bibnamefont
  {Malgaretti}}, \bibinfo {author} {\bibfnamefont {P.}~\bibnamefont
  {Nowakowski}}, \ and\ \bibinfo {author} {\bibfnamefont {H.}~\bibnamefont
  {Stark}},\ }\href {\doibase 10.1209/0295-5075/134/20002} {\bibfield
  {journal} {\bibinfo  {journal} {{EPL} (Europhysics Letters)}\ }\textbf
  {\bibinfo {volume} {134}},\ \bibinfo {pages} {20002} (\bibinfo {year}
  {2021})}\BibitemShut {NoStop}%
\bibitem [{\citenamefont {Fodor}\ and\ \citenamefont
  {Cates}(2021)}]{Fodor2021}%
  \BibitemOpen
  \bibfield  {author} {\bibinfo {author} {\bibfnamefont {{\'{E}}.}~\bibnamefont
  {Fodor}}\ and\ \bibinfo {author} {\bibfnamefont {M.~E.}\ \bibnamefont
  {Cates}},\ }\href {\doibase 10.1209/0295-5075/134/10003} {\bibfield
  {journal} {\bibinfo  {journal} {EPL}\ }\textbf {\bibinfo {volume} {134}},\
  \bibinfo {pages} {10003} (\bibinfo {year} {2021})}\BibitemShut {NoStop}%
\bibitem [{\citenamefont {Gronchi}\ and\ \citenamefont
  {Puglisi}(2021)}]{Gronchi2021}%
  \BibitemOpen
  \bibfield  {author} {\bibinfo {author} {\bibfnamefont {G.}~\bibnamefont
  {Gronchi}}\ and\ \bibinfo {author} {\bibfnamefont {A.}~\bibnamefont
  {Puglisi}},\ }\href {\doibase 10.1103/PhysRevE.103.052134} {\bibfield
  {journal} {\bibinfo  {journal} {Phys. Rev. E}\ }\textbf {\bibinfo {volume}
  {103}},\ \bibinfo {pages} {052134} (\bibinfo {year} {2021})}\BibitemShut
  {NoStop}%
\bibitem [{\citenamefont {Speck}(2022)}]{Speck2022}%
  \BibitemOpen
  \bibfield  {author} {\bibinfo {author} {\bibfnamefont {T.}~\bibnamefont
  {Speck}},\ }\href {\doibase 10.1103/PhysRevE.105.L012601} {\bibfield
  {journal} {\bibinfo  {journal} {Phys. Rev. E}\ }\textbf {\bibinfo {volume}
  {105}},\ \bibinfo {pages} {L012601} (\bibinfo {year} {2022})}\BibitemShut
  {NoStop}%
\bibitem [{\citenamefont {Ion}\ and\ \citenamefont {Urna}(2022)}]{Santra2022}%
  \BibitemOpen
  \bibfield  {author} {\bibinfo {author} {\bibfnamefont {S.}~\bibnamefont
  {Ion}}\ and\ \bibinfo {author} {\bibfnamefont {B.}~\bibnamefont {Urna}},\
  }\href {\doibase 10.48550/arXiv.2201.00796} {\bibfield  {journal} {\bibinfo
  {journal} {Arxiv}\ ,\ \bibinfo {pages} {2201.00796}} (\bibinfo {year}
  {2022})}\BibitemShut {NoStop}%
\bibitem [{\citenamefont {Sokolov}\ \emph {et~al.}(2010)\citenamefont
  {Sokolov}, \citenamefont {Apodaca}, \citenamefont {Grzybowski},\ and\
  \citenamefont {Aranson}}]{Sokolov2010}%
  \BibitemOpen
  \bibfield  {author} {\bibinfo {author} {\bibfnamefont {A.}~\bibnamefont
  {Sokolov}}, \bibinfo {author} {\bibfnamefont {M.~M.}\ \bibnamefont
  {Apodaca}}, \bibinfo {author} {\bibfnamefont {B.~A.}\ \bibnamefont
  {Grzybowski}}, \ and\ \bibinfo {author} {\bibfnamefont {I.~S.}\ \bibnamefont
  {Aranson}},\ }\href {\doibase 10.1073/pnas.0913015107} {\bibfield  {journal}
  {\bibinfo  {journal} {Proc. Natl. Acad. Sci.}\ }\textbf {\bibinfo {volume}
  {107}},\ \bibinfo {pages} {969} (\bibinfo {year} {2010})}\BibitemShut
  {NoStop}%
\bibitem [{\citenamefont {Di~Leonardo}\ \emph {et~al.}(2010)\citenamefont
  {Di~Leonardo}, \citenamefont {Angelani}, \citenamefont
  {Dell{\textquoteright}Arciprete}, \citenamefont {Ruocco}, \citenamefont
  {Iebba}, \citenamefont {Schippa}, \citenamefont {Conte}, \citenamefont
  {Mecarini}, \citenamefont {De~Angelis},\ and\ \citenamefont
  {Di~Fabrizio}}]{DiLeonardo2010}%
  \BibitemOpen
  \bibfield  {author} {\bibinfo {author} {\bibfnamefont {R.}~\bibnamefont
  {Di~Leonardo}}, \bibinfo {author} {\bibfnamefont {L.}~\bibnamefont
  {Angelani}}, \bibinfo {author} {\bibfnamefont {D.}~\bibnamefont
  {Dell{\textquoteright}Arciprete}}, \bibinfo {author} {\bibfnamefont
  {G.}~\bibnamefont {Ruocco}}, \bibinfo {author} {\bibfnamefont
  {V.}~\bibnamefont {Iebba}}, \bibinfo {author} {\bibfnamefont
  {S.}~\bibnamefont {Schippa}}, \bibinfo {author} {\bibfnamefont {M.~P.}\
  \bibnamefont {Conte}}, \bibinfo {author} {\bibfnamefont {F.}~\bibnamefont
  {Mecarini}}, \bibinfo {author} {\bibfnamefont {F.}~\bibnamefont
  {De~Angelis}}, \ and\ \bibinfo {author} {\bibfnamefont {E.}~\bibnamefont
  {Di~Fabrizio}},\ }\href {\doibase 10.1073/pnas.0910426107} {\bibfield
  {journal} {\bibinfo  {journal} {Proc. Natl. Acad. Sci.}\ }\textbf {\bibinfo
  {volume} {107}},\ \bibinfo {pages} {9541} (\bibinfo {year}
  {2010})}\BibitemShut {NoStop}%
\bibitem [{\citenamefont {Malgaretti}\ \emph {et~al.}(2013)\citenamefont
  {Malgaretti}, \citenamefont {Pagonabarraga},\ and\ \citenamefont
  {Rubi}}]{Malgaretti2013JCP}%
  \BibitemOpen
  \bibfield  {author} {\bibinfo {author} {\bibfnamefont {P.}~\bibnamefont
  {Malgaretti}}, \bibinfo {author} {\bibfnamefont {I.}~\bibnamefont
  {Pagonabarraga}}, \ and\ \bibinfo {author} {\bibfnamefont {J.~M.}\
  \bibnamefont {Rubi}},\ }\href {\doibase https://doi.org/10.1063/1.4804632}
  {\bibfield  {journal} {\bibinfo  {journal} {J. Chem. Phys.}\ }\textbf
  {\bibinfo {volume} {138}},\ \bibinfo {pages} {194906} (\bibinfo {year}
  {2013})}\BibitemShut {NoStop}%
\bibitem [{\citenamefont {Malgaretti}\ \emph {et~al.}(2014)\citenamefont
  {Malgaretti}, \citenamefont {Pagonabarraga},\ and\ \citenamefont
  {Rubi}}]{Malgaretti2014EPJ}%
  \BibitemOpen
  \bibfield  {author} {\bibinfo {author} {\bibfnamefont {P.}~\bibnamefont
  {Malgaretti}}, \bibinfo {author} {\bibfnamefont {I.}~\bibnamefont
  {Pagonabarraga}}, \ and\ \bibinfo {author} {\bibfnamefont {J.}~\bibnamefont
  {Rubi}},\ }\href {\doibase DOI: 10.1140/epjst/e2014-02334-4} {\bibfield
  {journal} {\bibinfo  {journal} {European Physics Journal Special Topics}\
  }\textbf {\bibinfo {volume} {223}},\ \bibinfo {pages} {3295} (\bibinfo {year}
  {2014})}\BibitemShut {NoStop}%
\bibitem [{\citenamefont {Yang}\ and\ \citenamefont
  {Ripoll}(2014)}]{Ripoll2014}%
  \BibitemOpen
  \bibfield  {author} {\bibinfo {author} {\bibfnamefont {M.}~\bibnamefont
  {Yang}}\ and\ \bibinfo {author} {\bibfnamefont {M.}~\bibnamefont {Ripoll}},\
  }\href {\doibase 10.1039/C3SM52417E} {\bibfield  {journal} {\bibinfo
  {journal} {Soft Matter}\ }\textbf {\bibinfo {volume} {10}},\ \bibinfo {pages}
  {1006} (\bibinfo {year} {2014})}\BibitemShut {NoStop}%
\bibitem [{\citenamefont {Michelin}\ \emph {et~al.}(2015)\citenamefont
  {Michelin}, \citenamefont {Montenegro-Johnson}, \citenamefont {De~Canio},
  \citenamefont {Lobato-Dauzier},\ and\ \citenamefont {Lauga}}]{Michelin2015}%
  \BibitemOpen
  \bibfield  {author} {\bibinfo {author} {\bibfnamefont {S.}~\bibnamefont
  {Michelin}}, \bibinfo {author} {\bibfnamefont {T.~D.}\ \bibnamefont
  {Montenegro-Johnson}}, \bibinfo {author} {\bibfnamefont {G.}~\bibnamefont
  {De~Canio}}, \bibinfo {author} {\bibfnamefont {N.}~\bibnamefont
  {Lobato-Dauzier}}, \ and\ \bibinfo {author} {\bibfnamefont {E.}~\bibnamefont
  {Lauga}},\ }\href {\doibase 10.1039/C5SM00718F} {\bibfield  {journal}
  {\bibinfo  {journal} {Soft Matter}\ }\textbf {\bibinfo {volume} {11}},\
  \bibinfo {pages} {5804} (\bibinfo {year} {2015})}\BibitemShut {NoStop}%
\bibitem [{\citenamefont {Malgaretti}\ and\ \citenamefont
  {Stark}(2017)}]{Malgaretti2017}%
  \BibitemOpen
  \bibfield  {author} {\bibinfo {author} {\bibfnamefont {P.}~\bibnamefont
  {Malgaretti}}\ and\ \bibinfo {author} {\bibfnamefont {H.}~\bibnamefont
  {Stark}},\ }\href {\doibase 10.1063/1.4981886} {\bibfield  {journal}
  {\bibinfo  {journal} {J. Chem. Phys.}\ }\textbf {\bibinfo {volume} {146}},\
  \bibinfo {pages} {174901} (\bibinfo {year} {2017})}\BibitemShut {NoStop}%
\bibitem [{\citenamefont {Holubec}\ \emph {et~al.}(2017)\citenamefont
  {Holubec}, \citenamefont {Ryabov}, \citenamefont {Yaghoubi}, \citenamefont
  {Varga}, \citenamefont {Khodaee}, \citenamefont {Foulaadvand},\ and\
  \citenamefont {Chvosta}}]{Ryabov2017}%
  \BibitemOpen
  \bibfield  {author} {\bibinfo {author} {\bibfnamefont {V.}~\bibnamefont
  {Holubec}}, \bibinfo {author} {\bibfnamefont {A.}~\bibnamefont {Ryabov}},
  \bibinfo {author} {\bibfnamefont {M.~H.}\ \bibnamefont {Yaghoubi}}, \bibinfo
  {author} {\bibfnamefont {M.}~\bibnamefont {Varga}}, \bibinfo {author}
  {\bibfnamefont {A.}~\bibnamefont {Khodaee}}, \bibinfo {author} {\bibfnamefont
  {M.~E.}\ \bibnamefont {Foulaadvand}}, \ and\ \bibinfo {author} {\bibfnamefont
  {P.}~\bibnamefont {Chvosta}},\ }\href {\doibase 10.3390/e19040119} {\bibfield
   {journal} {\bibinfo  {journal} {Entropy}\ }\textbf {\bibinfo {volume} {19}}
  (\bibinfo {year} {2017}),\ 10.3390/e19040119}\BibitemShut {NoStop}%
\bibitem [{\citenamefont {Zanin}\ \emph {et~al.}(2021)\citenamefont {Zanin},
  \citenamefont {Antesberger}, \citenamefont {Jacquet}, \citenamefont
  {Ribeiro}, \citenamefont {Rozema},\ and\ \citenamefont
  {Walther}}]{Zanin2021}%
  \BibitemOpen
  \bibfield  {author} {\bibinfo {author} {\bibfnamefont {G.~L.}\ \bibnamefont
  {Zanin}}, \bibinfo {author} {\bibfnamefont {M.}~\bibnamefont {Antesberger}},
  \bibinfo {author} {\bibfnamefont {M.~J.}\ \bibnamefont {Jacquet}}, \bibinfo
  {author} {\bibfnamefont {P.~H.~S.}\ \bibnamefont {Ribeiro}}, \bibinfo
  {author} {\bibfnamefont {L.~A.}\ \bibnamefont {Rozema}}, \ and\ \bibinfo
  {author} {\bibfnamefont {P.}~\bibnamefont {Walther}},\ }\href@noop {} {\
  (\bibinfo {year} {2021})},\ \Eprint {http://arxiv.org/abs/2107.09686}
  {arXiv:2107.09686 [quant-ph]} \BibitemShut {NoStop}%
\bibitem [{\citenamefont {Marini Bettolo~Marconi}\ and\ \citenamefont
  {Maggi}(2015)}]{Marconi2015}%
  \BibitemOpen
  \bibfield  {author} {\bibinfo {author} {\bibfnamefont {U.}~\bibnamefont
  {Marini Bettolo~Marconi}}\ and\ \bibinfo {author} {\bibfnamefont
  {C.}~\bibnamefont {Maggi}},\ }\href {\doibase 10.1039/C5SM01718A} {\bibfield
  {journal} {\bibinfo  {journal} {Soft Matter}\ }\textbf {\bibinfo {volume}
  {11}},\ \bibinfo {pages} {8768} (\bibinfo {year} {2015})}\BibitemShut
  {NoStop}%
\bibitem [{\citenamefont {Landauer}(1961)}]{Landauer1961}%
  \BibitemOpen
  \bibfield  {author} {\bibinfo {author} {\bibfnamefont {R.}~\bibnamefont
  {Landauer}},\ }\href {\doibase 10.1147/rd.53.0183} {\bibfield  {journal}
  {\bibinfo  {journal} {IBM Journal of Research and Development}\ }\textbf
  {\bibinfo {volume} {5}},\ \bibinfo {pages} {183} (\bibinfo {year}
  {1961})}\BibitemShut {NoStop}%
\bibitem [{\citenamefont {B{\'e}rut}\ \emph {et~al.}(2012)\citenamefont
  {B{\'e}rut}, \citenamefont {Arakelyan}, \citenamefont {Petrosyan},
  \citenamefont {Ciliberto}, \citenamefont {Dillenschneider},\ and\
  \citenamefont {Lutz}}]{Berut2012}%
  \BibitemOpen
  \bibfield  {author} {\bibinfo {author} {\bibfnamefont {A.}~\bibnamefont
  {B{\'e}rut}}, \bibinfo {author} {\bibfnamefont {A.}~\bibnamefont
  {Arakelyan}}, \bibinfo {author} {\bibfnamefont {A.}~\bibnamefont
  {Petrosyan}}, \bibinfo {author} {\bibfnamefont {S.}~\bibnamefont
  {Ciliberto}}, \bibinfo {author} {\bibfnamefont {R.}~\bibnamefont
  {Dillenschneider}}, \ and\ \bibinfo {author} {\bibfnamefont {E.}~\bibnamefont
  {Lutz}},\ }\href {\doibase 10.1038/nature10872} {\bibfield  {journal}
  {\bibinfo  {journal} {Nature}\ }\textbf {\bibinfo {volume} {483}},\ \bibinfo
  {pages} {187} (\bibinfo {year} {2012})}\BibitemShut {NoStop}%
\bibitem [{\citenamefont {Kawai}\ \emph {et~al.}(2007)\citenamefont {Kawai},
  \citenamefont {Parrondo},\ and\ \citenamefont {den Broeck}}]{VdBroek2007}%
  \BibitemOpen
  \bibfield  {author} {\bibinfo {author} {\bibfnamefont {R.}~\bibnamefont
  {Kawai}}, \bibinfo {author} {\bibfnamefont {J.~M.~R.}\ \bibnamefont
  {Parrondo}}, \ and\ \bibinfo {author} {\bibfnamefont {C.~V.}\ \bibnamefont
  {den Broeck}},\ }\href {\doibase 10.1103/PhysRevLett.98.080602} {\bibfield
  {journal} {\bibinfo  {journal} {Phys. Rev. Lett.}\ }\textbf {\bibinfo
  {volume} {98}},\ \bibinfo {pages} {080602} (\bibinfo {year}
  {2007})}\BibitemShut {NoStop}%
\bibitem [{\citenamefont {Schmiedl}\ and\ \citenamefont
  {Seifert}(2007)}]{Schmiedl2007}%
  \BibitemOpen
  \bibfield  {author} {\bibinfo {author} {\bibfnamefont {T.}~\bibnamefont
  {Schmiedl}}\ and\ \bibinfo {author} {\bibfnamefont {U.}~\bibnamefont
  {Seifert}},\ }\href {\doibase 10.1209/0295-5075/81/20003} {\bibfield
  {journal} {\bibinfo  {journal} {{EPL} (Europhys. Lett.)}\ }\textbf {\bibinfo
  {volume} {81}},\ \bibinfo {pages} {20003} (\bibinfo {year}
  {2007})}\BibitemShut {NoStop}%
\bibitem [{\citenamefont {Kjelstrup}\ \emph {et~al.}(2010)\citenamefont
  {Kjelstrup}, \citenamefont {Bedeaux}, \citenamefont {Johannessen},\ and\
  \citenamefont {Gross}}]{Kajelstrup_book}%
  \BibitemOpen
  \bibfield  {author} {\bibinfo {author} {\bibfnamefont {S.}~\bibnamefont
  {Kjelstrup}}, \bibinfo {author} {\bibfnamefont {D.}~\bibnamefont {Bedeaux}},
  \bibinfo {author} {\bibfnamefont {E.}~\bibnamefont {Johannessen}}, \ and\
  \bibinfo {author} {\bibfnamefont {J.}~\bibnamefont {Gross}},\ }\href@noop {}
  {\emph {\bibinfo {title} {Non-Equilibrium Thermodynamics for Engineers}}}\
  (\bibinfo  {publisher} {World Scientific Publishing Company, Singapore},\
  \bibinfo {year} {2010})\BibitemShut {NoStop}%
\bibitem [{\citenamefont {Sabass}\ and\ \citenamefont
  {Seifert}(2012)}]{Benedikt2012}%
  \BibitemOpen
  \bibfield  {author} {\bibinfo {author} {\bibfnamefont {B.}~\bibnamefont
  {Sabass}}\ and\ \bibinfo {author} {\bibfnamefont {U.}~\bibnamefont
  {Seifert}},\ }\href {\doibase 10.1063/1.3681143} {\bibfield  {journal}
  {\bibinfo  {journal} {The Journal of Chemical Physics}\ }\textbf {\bibinfo
  {volume} {136}},\ \bibinfo {pages} {064508} (\bibinfo {year}
  {2012})}\BibitemShut {NoStop}%
\bibitem [{\citenamefont {Shah}\ \emph {et~al.}(2020)\citenamefont {Shah},
  \citenamefont {Wang}, \citenamefont {Xian}, \citenamefont {Zhou},
  \citenamefont {Chen}, \citenamefont {Lin},\ and\ \citenamefont
  {Gao}}]{Shah2020}%
  \BibitemOpen
  \bibfield  {author} {\bibinfo {author} {\bibfnamefont {Z.~H.}\ \bibnamefont
  {Shah}}, \bibinfo {author} {\bibfnamefont {S.}~\bibnamefont {Wang}}, \bibinfo
  {author} {\bibfnamefont {L.}~\bibnamefont {Xian}}, \bibinfo {author}
  {\bibfnamefont {X.}~\bibnamefont {Zhou}}, \bibinfo {author} {\bibfnamefont
  {Y.}~\bibnamefont {Chen}}, \bibinfo {author} {\bibfnamefont {G.}~\bibnamefont
  {Lin}}, \ and\ \bibinfo {author} {\bibfnamefont {Y.}~\bibnamefont {Gao}},\
  }\href {\doibase 10.1039/D0CC06812H} {\bibfield  {journal} {\bibinfo
  {journal} {Chem. Commun.}\ }\textbf {\bibinfo {volume} {56}},\ \bibinfo
  {pages} {15301} (\bibinfo {year} {2020})}\BibitemShut {NoStop}%
\bibitem [{\citenamefont {Friedrich}(2018)}]{Friederich2018}%
  \BibitemOpen
  \bibfield  {author} {\bibinfo {author} {\bibfnamefont {B.~M.}\ \bibnamefont
  {Friedrich}},\ }\href {\doibase 10.1103/PhysRevE.97.042416} {\bibfield
  {journal} {\bibinfo  {journal} {Phys. Rev. E}\ }\textbf {\bibinfo {volume}
  {97}},\ \bibinfo {pages} {042416} (\bibinfo {year} {2018})}\BibitemShut
  {NoStop}%
\bibitem [{\citenamefont {Katsu-Kimura}\ \emph {et~al.}(2009)\citenamefont
  {Katsu-Kimura}, \citenamefont {Nakaya}, \citenamefont {Baba},\ and\
  \citenamefont {Mogami}}]{Katsu-Kimura2009}%
  \BibitemOpen
  \bibfield  {author} {\bibinfo {author} {\bibfnamefont {Y.}~\bibnamefont
  {Katsu-Kimura}}, \bibinfo {author} {\bibfnamefont {F.}~\bibnamefont
  {Nakaya}}, \bibinfo {author} {\bibfnamefont {S.~A.}\ \bibnamefont {Baba}}, \
  and\ \bibinfo {author} {\bibfnamefont {Y.}~\bibnamefont {Mogami}},\ }\href
  {\doibase 10.1242/jeb.028894} {\bibfield  {journal} {\bibinfo  {journal}
  {Journal of Experimental Biology}\ }\textbf {\bibinfo {volume} {212}},\
  \bibinfo {pages} {1819} (\bibinfo {year} {2009})}\BibitemShut {NoStop}%
\bibitem [{\citenamefont {Chen}\ \emph {et~al.}(2015)\citenamefont {Chen},
  \citenamefont {Heymann}, \citenamefont {Fraden}, \citenamefont {Nicastro},\
  and\ \citenamefont {Dogic}}]{Daniel2015}%
  \BibitemOpen
  \bibfield  {author} {\bibinfo {author} {\bibfnamefont {D.~T.}\ \bibnamefont
  {Chen}}, \bibinfo {author} {\bibfnamefont {M.}~\bibnamefont {Heymann}},
  \bibinfo {author} {\bibfnamefont {S.}~\bibnamefont {Fraden}}, \bibinfo
  {author} {\bibfnamefont {D.}~\bibnamefont {Nicastro}}, \ and\ \bibinfo
  {author} {\bibfnamefont {Z.}~\bibnamefont {Dogic}},\ }\href {\doibase
  https://doi.org/10.1016/j.bpj.2015.11.003} {\bibfield  {journal} {\bibinfo
  {journal} {Biophysical Journal}\ }\textbf {\bibinfo {volume} {109}},\
  \bibinfo {pages} {2562} (\bibinfo {year} {2015})}\BibitemShut {NoStop}%
\bibitem [{\citenamefont {Maggi}\ \emph {et~al.}(2015)\citenamefont {Maggi},
  \citenamefont {Saglimbeni}, \citenamefont {Dipalo}, \citenamefont
  {De~Angelis},\ and\ \citenamefont {Di~Leonardo}}]{Maggi2015}%
  \BibitemOpen
  \bibfield  {author} {\bibinfo {author} {\bibfnamefont {C.}~\bibnamefont
  {Maggi}}, \bibinfo {author} {\bibfnamefont {F.}~\bibnamefont {Saglimbeni}},
  \bibinfo {author} {\bibfnamefont {M.}~\bibnamefont {Dipalo}}, \bibinfo
  {author} {\bibfnamefont {F.}~\bibnamefont {De~Angelis}}, \ and\ \bibinfo
  {author} {\bibfnamefont {R.}~\bibnamefont {Di~Leonardo}},\ }\href {\doibase
  10.1038/ncomms8855} {\bibfield  {journal} {\bibinfo  {journal} {Nature
  Communications}\ }\textbf {\bibinfo {volume} {6}},\ \bibinfo {pages} {7855}
  (\bibinfo {year} {2015})}\BibitemShut {NoStop}%
\bibitem [{\citenamefont {Tobias~B\"auerle}\ and\ \citenamefont
  {Bechinger}(2020)}]{Baeuerle2020}%
  \BibitemOpen
  \bibfield  {author} {\bibinfo {author} {\bibfnamefont {R.~C.~L.}\
  \bibnamefont {Tobias~B\"auerle}}\ and\ \bibinfo {author} {\bibfnamefont
  {C.}~\bibnamefont {Bechinger}},\ }\href@noop {} {\ ,\ \bibinfo {pages} {2547}
  (\bibinfo {year} {2020})}\BibitemShut {NoStop}%
\end{thebibliography}%
\newpage

\onecolumngrid

\section{Supplemental Material}

\pa{
\subsection{Active Brownian Particle (ABP)}\label{app:ABP}
The motion of an active Brownian  particle is governed by (see Ref.~\cite{RomanczukSchimansky-Geier2012,Zoettl2016})
\begin{align}
\gamma_t \dot{\mathbf{x}}(t)=\,v_0 \boldsymbol{\theta}(t)+\boldsymbol{\xi}(t)\,,\quad \dot{\boldsymbol{\theta}}=\boldsymbol{\eta}(t)\times \boldsymbol{\theta}
\label{eq:langevin}
\end{align}
where, $v_0$ is the magnitude of the active velocity, $\boldsymbol{\theta}$ its direction and, introducing the translational and rotational diffusion coefficients  $D_{t,r}$, $\boldsymbol{\eta}$ and $\boldsymbol{\xi}$ are delta-correlated white noise
\begin{subequations}\label{eq:noise}
\begin{align}
\langle \boldsymbol{\xi}(t) \rangle = & 0\,,\quad \langle  \boldsymbol{\xi}(t)\boldsymbol{\xi}(t')\rangle = 2 D_t \delta(t-t')\\
\langle \boldsymbol{\eta}(t) \rangle = & 0\,,\quad \langle  \boldsymbol{\eta}(t)\boldsymbol{\eta}(t')\rangle = 2 D_r \delta(t-t')
\end{align}
\end{subequations}
This model leads to a finite correlation time of the active component of velocity (see Ref.~\cite{RomanczukSchimansky-Geier2012,Zoettl2016})
\begin{align}
v_0^2 \langle \boldsymbol{\theta}(t) \cdot \boldsymbol{\theta}(t')\rangle = v_0^2 e^{-\frac{|t-t'|}{\tau_M}}
\end{align}
with $\tau_M = \frac{1}{2D_r}$.
}

\pa{
\subsection{Alternative realization of the \textit{quasi-static} active Szilard engine}
\begin{figure}[h]
\centering
\includegraphics[scale=0.45,angle=0]{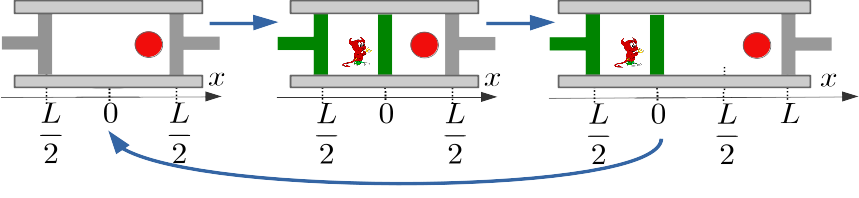}
\caption{Cartoon of an alternative realization of the \textit{quasi-static} active Szilard engine. 
The daemon divides the cylinder by an additional wall (in green) and at the same time fixes the piston (in green) of the empty part of the cylinder. After the expansion of the right piston, the daemon removes the wall, let the system compress to the original size, and then releases the left piston.}
\label{fig:scheme0}
\end{figure}
The Maxwell daemon in this type of Szilard engine operates as follows. A single particle confined between two pistons of area $A$ exerts a pressure $\beta\Pi_0$ on the pistons such that the stationary distance between them is $L$\footnote{For a single ideal gas particle $\beta\Pi_0=(AL)^{-1}$.} (see Fig.\ref{fig:scheme0}). 
 The daemon detects the position of the particle, inserts a wall in the middle of the box ($x=0$), and at the same time fixes the position of the piston of the empty chamber (depicted in green in Fig.\ref{fig:scheme0}). Due to the reduction of volume available to the particle, the pressure in the  occupied chamber increases and the volume of the chamber expands till $x=L$, i.e., when the pressure matches the outer pressure $\Pi_0$. Once $x=L$ is reached, the expansion stops since, by 
definition, $\Pi=\Pi_0$. Then, the wall is released and the piston returns to $L/2$.
The work done by this engine during the expansion mathces $W_{exp}$ as reported in the main text. However, in the current case a (negative) work is performed during compression. 
Accordingly, using Eq.~\eqref{eq:Pi} we calculate the work during compression
\begin{align}
 \beta & W_{comp}=A \int_{\frac{3}{2}L}^{L}\!\!\! \beta \Pi\left(x\right) dx=\beta W_0\ln\left[\frac{1+\chi L}{1+3/2\chi L}\right]\,.
\end{align}
\begin{figure}
\includegraphics[scale=0.5,valign=t]{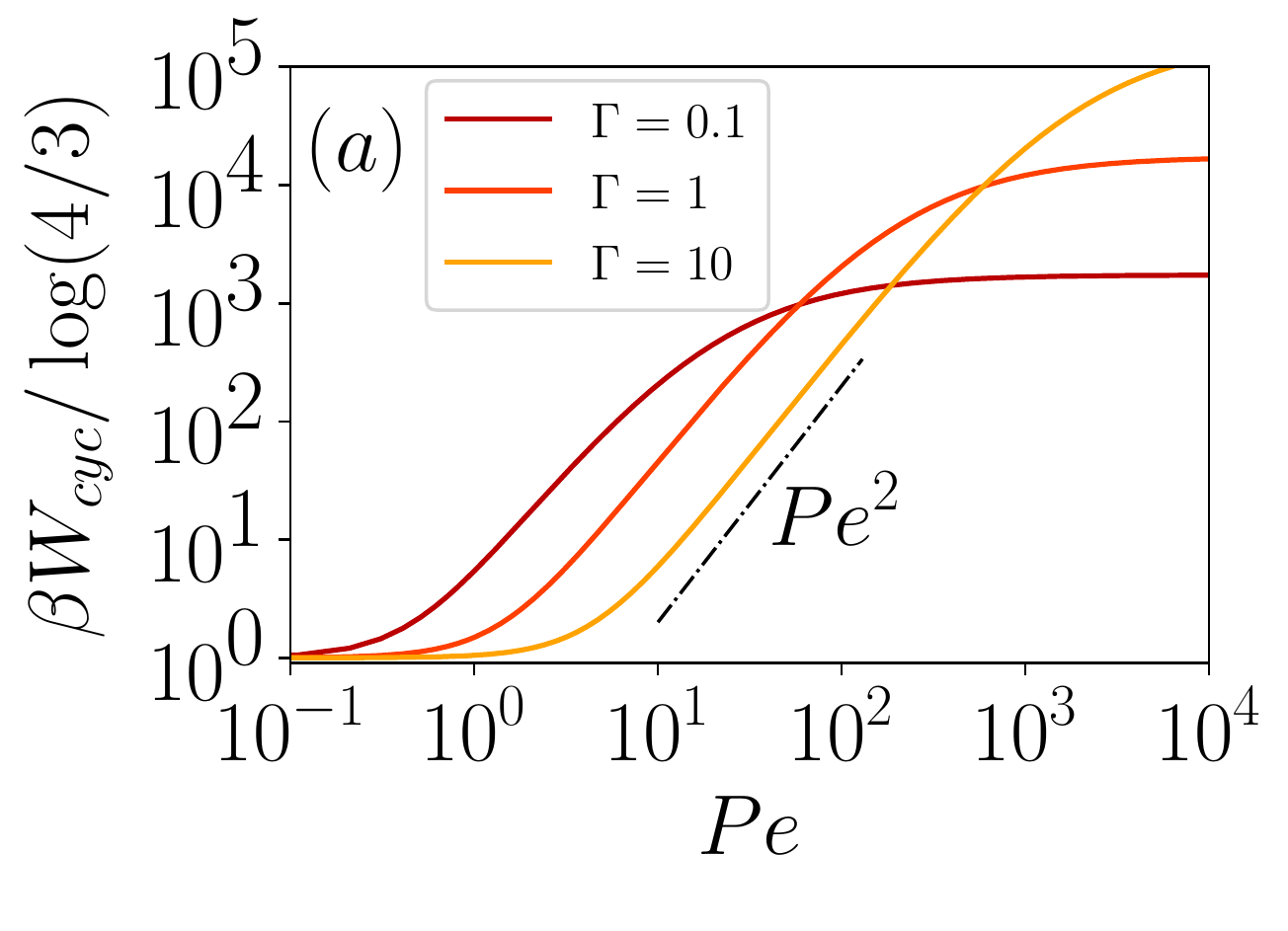}
\vspace{-5pt}\includegraphics[scale=0.5,valign=t]{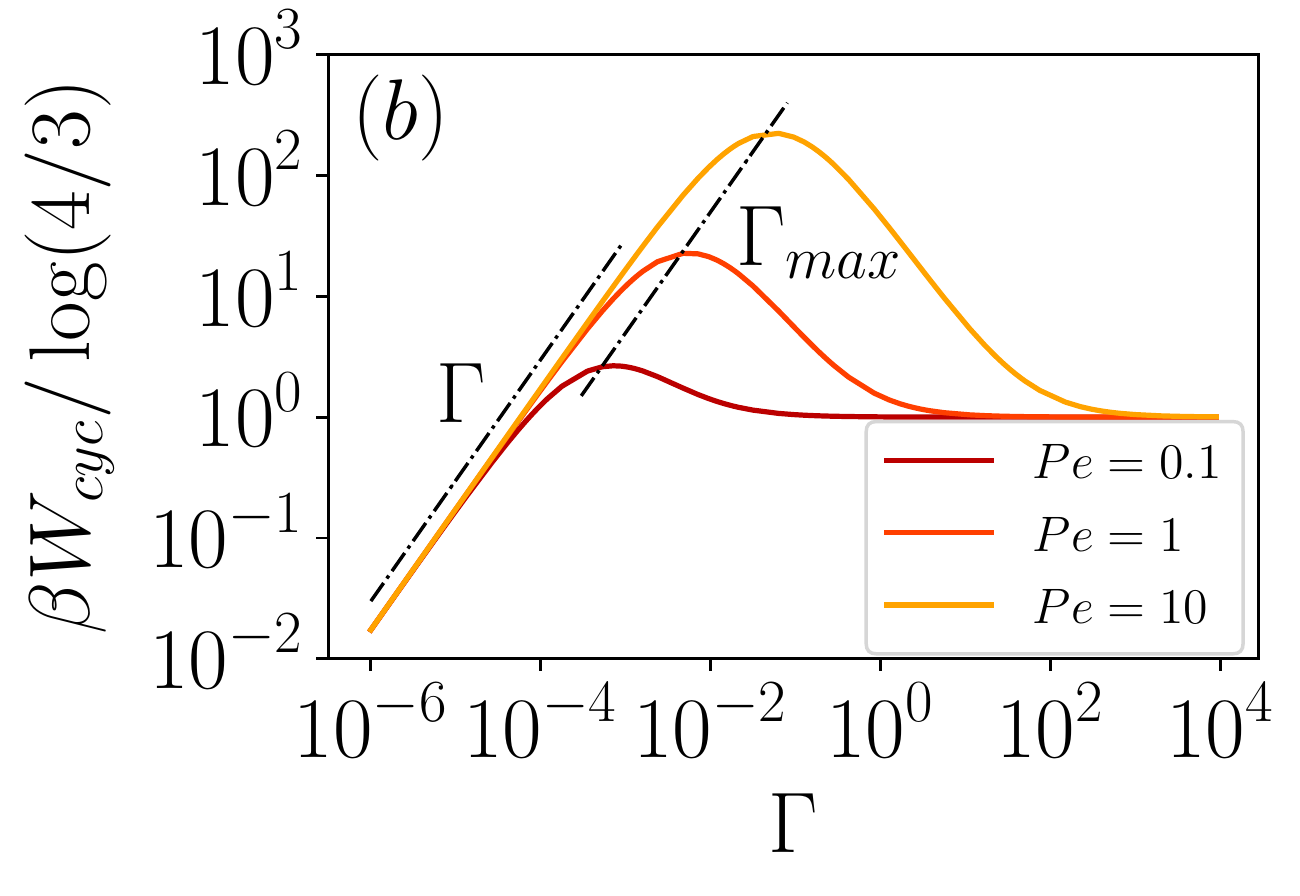}
\caption{Work extracted during a work cycle, $W_{cyc}$, normalized by the 
 value for thermal equilibrium, as function of Pe (upper panel) or $\Gamma$ (lower panel) with $L/R=100$.
\label{fig:W_cyc}}
\end{figure}
We remark that during compression the internal wall has been removed and hence the pressure of the ABP should be calculated taking into account that now the volume available to the particle is $A(x+L/2)$.
Accordingly, the work over a cycle, $\beta W_{cyc}=\beta(W_{exp}+W_{comp})$, reads:
\begin{align}
    \beta W_{cyc}=-\beta W_0\ln\left[1-\frac{1}{4}\left(\frac{\chi L}{1+\chi L}\right)^2\right]\,.
    \label{eq:W_cyc}
\end{align}
As shown in Fig.\ \ref{fig:W_cyc}, the work extracted over a cycle attains 
a plateau at $W_{cyc}  \simeq \ln(4/3)$ for small values of the P\'eclet number, $\Pe<\sqrt{2\Gamma}$,
and at $W_{cyc} \simeq \frac{\Gamma}{2}\frac{L^2 }{R^2}$ for large values,  $\Pe>2\Gamma L/R$.
For intermediate values of $\Pe$, $W_{cyc}$ scales as $\Pe^2$.
Interestingly, the extracted work has a non monotonous dependence on $\Gamma$  and $W_{cyc}$ is maximized for $\Gamma_{max}\sim \Pe$ (see Fig.S1 in Suppl. Mat.). Finally, for very large values of $\Gamma$, the active bath reduces to a thermal one, and the extracted work reaches a plateau $\beta W_{cyc} \simeq \ln(4/3)$ i.e., the work extracted from a thermal bath.\\
Finally, we report the values of $\Gamma_{max}$ obtained by the numerical solution of $\partial_{\Gamma} \beta W_{cyc}=0$ where $\beta W_{cyc}$ is defined in Eq.~(8) of the main text.
\begin{figure}[h!]
\includegraphics[scale=0.5]{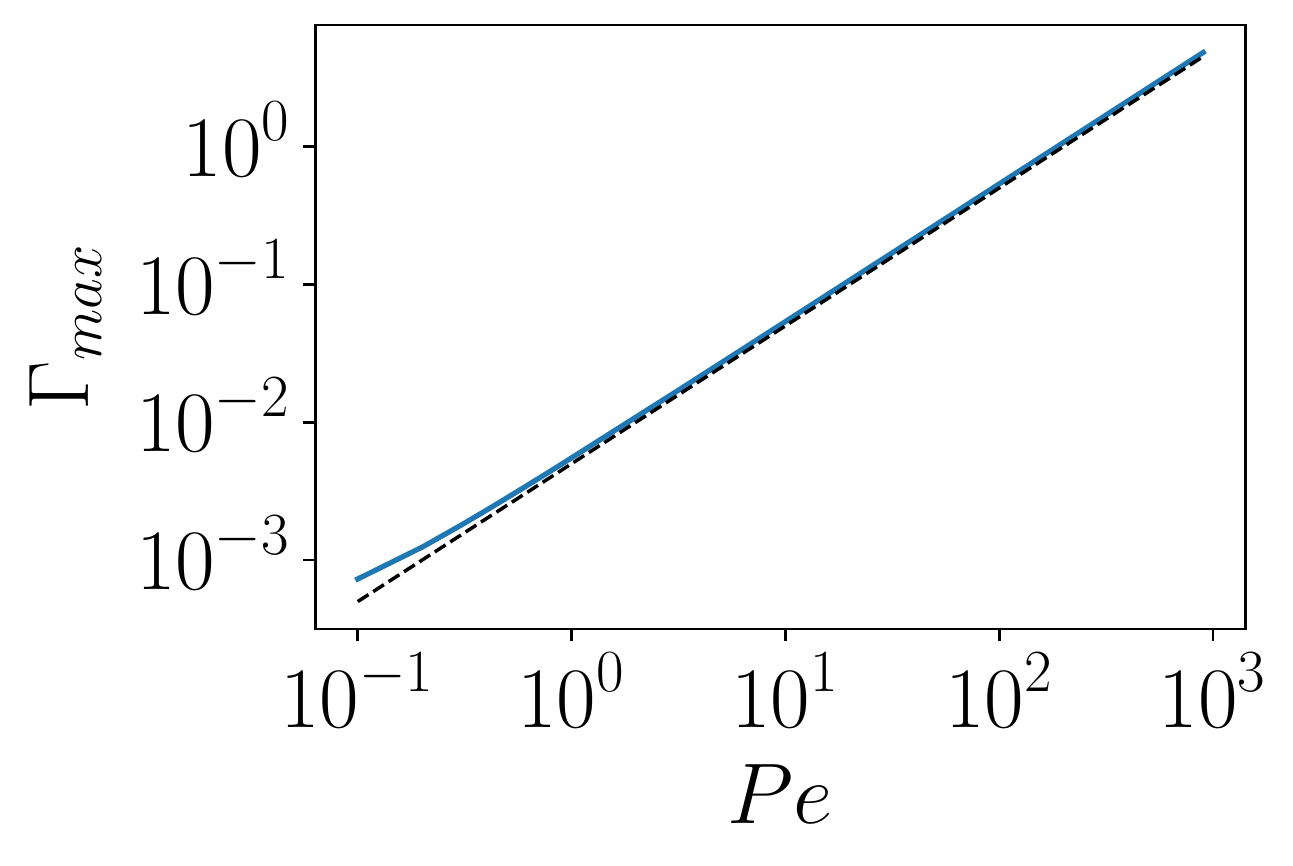}
\caption{
$\Gamma_{max}$, as function of $\Pe$.
\label{fig:Gamma-max}}
\end{figure}
It is interesting to note that within this realization of the Szilard engine, the dependence of  the extracted work on $\Gamma$ in non-monotonous, in contrast to the monotonic decay shown by the Szilard engine reported in the main text.}

\subsection{Derivation of Eq.~(17) of the main text}\label{app:W}
In the following we report the derivation of Eq.~(17). When the ABP is pushing against the wall its nominal velocity is reduced to 
\begin{align}
v_w = \frac{\gamma_p v_0-F}{\gamma_p + \gamma_w} = v_0 \frac{1-F/\gamma_p v_0}{1 + \gamma_w/\gamma_p} \equiv v_0\bar{v}_w
\label{eq:app_v_w}
\end{align}
where $\gamma_p$ is the friction coefficient of the ABP. Accordingly the work reads
\begin{align}
\left\langle W_{act}\right\rangle &=F \int_{\pa{\tau_{\delta}}}^\tau \left\langle v(t)\big |_{v(0)}\right\rangle dt\\
&=\frac{F}{v(0)}\int_{\pa{\tau_{\delta}}}^\tau\!\!\left\langle v(t)v(0)\right\rangle dt \\
&=\frac{F}{v_w}\int_{\pa{\tau_{\delta}}}^\tau\!\!\left\langle v(t)v(0)\right\rangle dt & \text{Using the definition of } v(0)\\
&=F\left(v_w+\frac{D}{v_w\tau_M}\right) \int_{\pa{\tau_{\delta}}}^\tau e^{-\frac{t}{\tau_M}} dt & \text{Using Eq.~(1)}\\
&= F\left(v_w+\frac{D}{v_w\tau_M}\right) \tau_M\left(\pa{e^{-\frac{\tau_\delta}{\tau_M}}}-e^{-\frac{\tau}{\tau_M}}\right)  & \text{integrating} \\
&= F v_w\tau_M \left(1+\frac{D}{v_w^2\tau_M}\right) \left(\pa{e^{-\frac{\tau_\delta}{\tau_M}}}-e^{-\frac{\tau}{\tau_M}}\right) & \text{grouping }v_w
\end{align}
Finally, for large activity $\frac{D}{v_w^2 \tau_M}\ll 1$ \pa{and high measurement precision $\tau_\delta \ll \min(\tau_M,\tau)$} the last expression reduces to
\begin{align}
\left\langle W_{act}\right\rangle\! = F v_w\tau_M \left(1-e^{-\bar{\tau}}\right)
\end{align}
where we have introduced the time ratio $\bar\tau=\tau/\tau_M$.
For later use we introduce the the rescaled work
\begin{align}
\overline{W}_{\!act} = \frac{\left\langle W_{act}\right\rangle}{\gamma_p v_0^2 \tau_M}=\frac{F v_w}{\gamma_p v_0^2}\left(1-e^{-\bar\tau}\right)=\bar{F}\bar{v}_w \left(1-e^{-\bar{\tau}}\right)
\label{eq:app_W_act}
\end{align}
where we introduced 
the dimensionless external force, $\bar{F}=F/v_0\gamma_p$ and we used Eq.~\eqref{eq:app_v_w}.
Then, we have
\begin{align}
\left\langle W_{act}\right\rangle\! = \gamma_p v_0^2 \tau_M \overline{W}_{\!act}\,.
\label{app:eq:W}
\end{align}

\subsection{Derivation of Eq.~(20) of the main text}\label{app:G}
Within a mean-field approximation, the energy dissipated by the piston during the time $\tau$ can be expressed as:
\begin{align}
\langle W^{pst}_{diss}\rangle &= \gamma_w \int_{\pa{\tau_\delta}}^\tau \!\! \left\langle v(t)\big |_{v(0)}\right\rangle^2  dt \\
& =\frac{\gamma_w}{v^2(0)}\int_{\pa{\tau_\delta}}^\tau\!\!\left\langle v(t)v(0)\right\rangle^2 dt \\
\end{align}
In the large activity regime we can approximate $v(0)=v_w^2+D/\tau_M\sim v_w^2$. Accordingly the last expression reduces to
\begin{align}
\langle W^{pst}_{diss}\rangle &=\gamma_w v_w^2 \int_{\pa{\tau_\delta}}^\tau e^{-2\frac{t}{\tau_M}} dt\\
& = \frac{\tau_M}{2} \gamma_w v_w^2 \left(\pa{e^{-2\frac{\tau_\delta}{\tau_M}}}- e^{-2\frac{\tau}{\tau_M}}\right)
\end{align}
\pa{As for the work, in the case of high measurement precision $\tau_\delta \ll \min(\tau_M,\tau)$ the previous expression reduces to 
\begin{align}
\langle W^{pst}_{diss}\rangle = \frac{\tau_M}{2} \gamma_w v_w^2 \left(1- e^{-2\frac{\tau}{\tau_M}}\right)
\end{align}
and} the overall dissipated energy hence reads:
\begin{align}
\langle W_{diss}\rangle &= \mathcal{P} \tau+\mathcal{M}+\frac{\tau_M}{2} \gamma_w v_w^2 \left(1- e^{-2\frac{\tau}{\tau_M}}\right)\nonumber\\
&\simeq \mathcal{P} \tau_M\left[\bar\tau+\overline{\mathcal{M}}+\frac{1}{2}\alpha \bar{\gamma}    \bar{v}_w^2\overline{W}^{pst}_{\!diss}\right]
\end{align}
where we used 
\begin{align}
\bar\gamma=&\frac{\gamma_w}{\gamma_p}\\
\overline{\mathcal{M}}=&\frac{\mathcal{M}}{\mathcal{P}\tau_M}\\
\frac{\gamma_w v_0^2}{\mathcal{P}}=\frac{\gamma_p v_0^2}{\mathcal{P}}\bar{\gamma}\simeq &\alpha \bar{\gamma}\\
\overline{W}^{pst}_{\!diss} =& 1- e^{-2\bar{\tau}}
\end{align}
Accordingly, the efficiency, Eq.~(20), reads
\begin{align}
\langle \eta\rangle &= \dfrac{1}{1+\dfrac{\langle W_{diss}\rangle}{\langle W_{act}\rangle}}\nonumber\\
&\simeq\dfrac{1}{1+\dfrac{1}{\alpha}\dfrac{\bar\tau+\overline{\mathcal{M}}+\frac{1}{2} \alpha \bar{\gamma}  \bar{v}_w^2\overline{W}^{pst}_{\!diss}}{\overline{W}_{\!act}}}
\label{eq:app_eta-tilde}
\end{align}
Finally
that is Eq.~(20) of the main text.

\subsection{Magnitude of $\frac{\gamma_w}{\gamma_p} \bar{v}_w^2\overline{W}^{pst}_{\!diss}$}\label{app:diss_pst}
\begin{figure}[h!]
\includegraphics[scale=0.5]{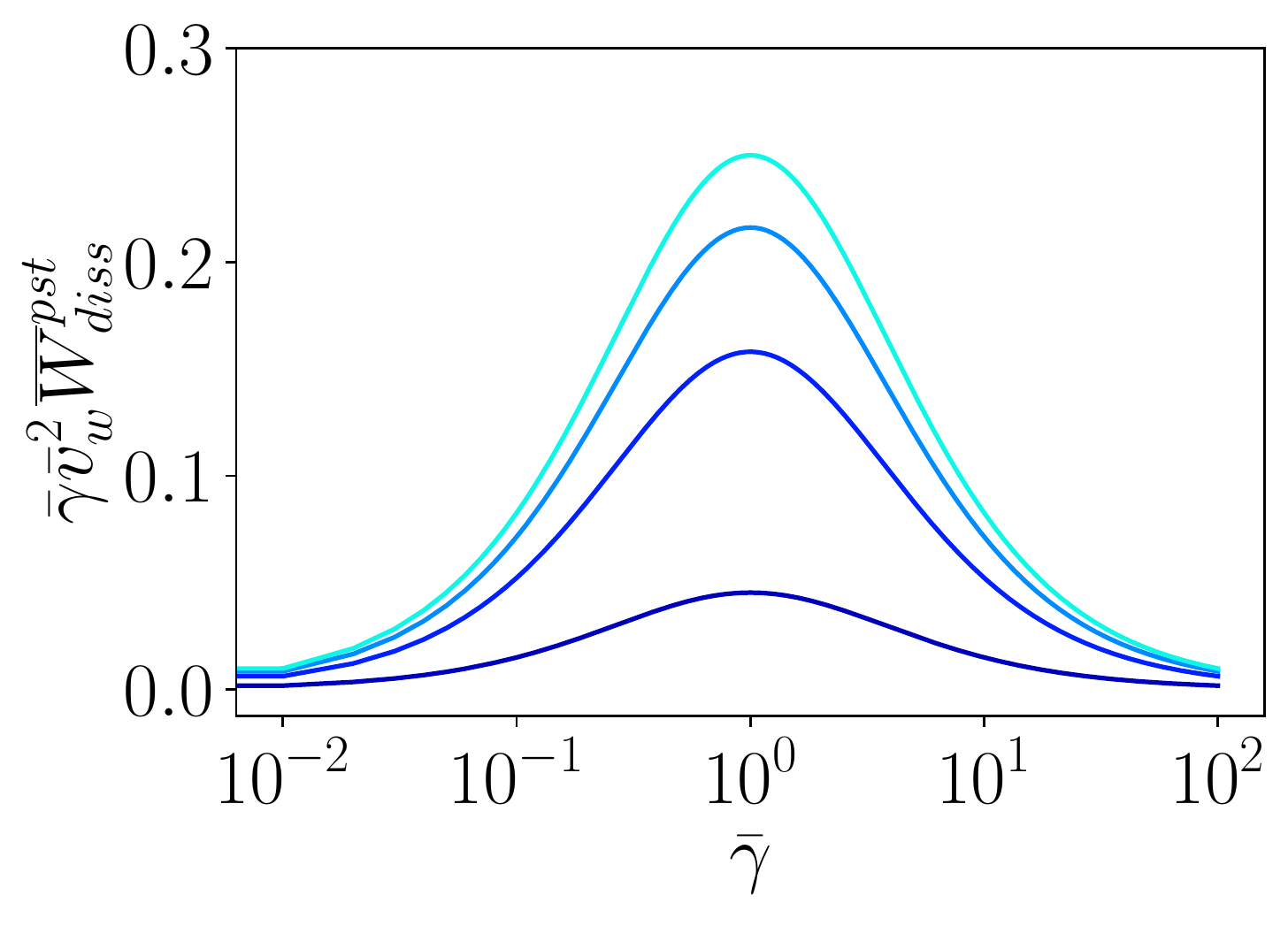}
\includegraphics[scale=0.5]{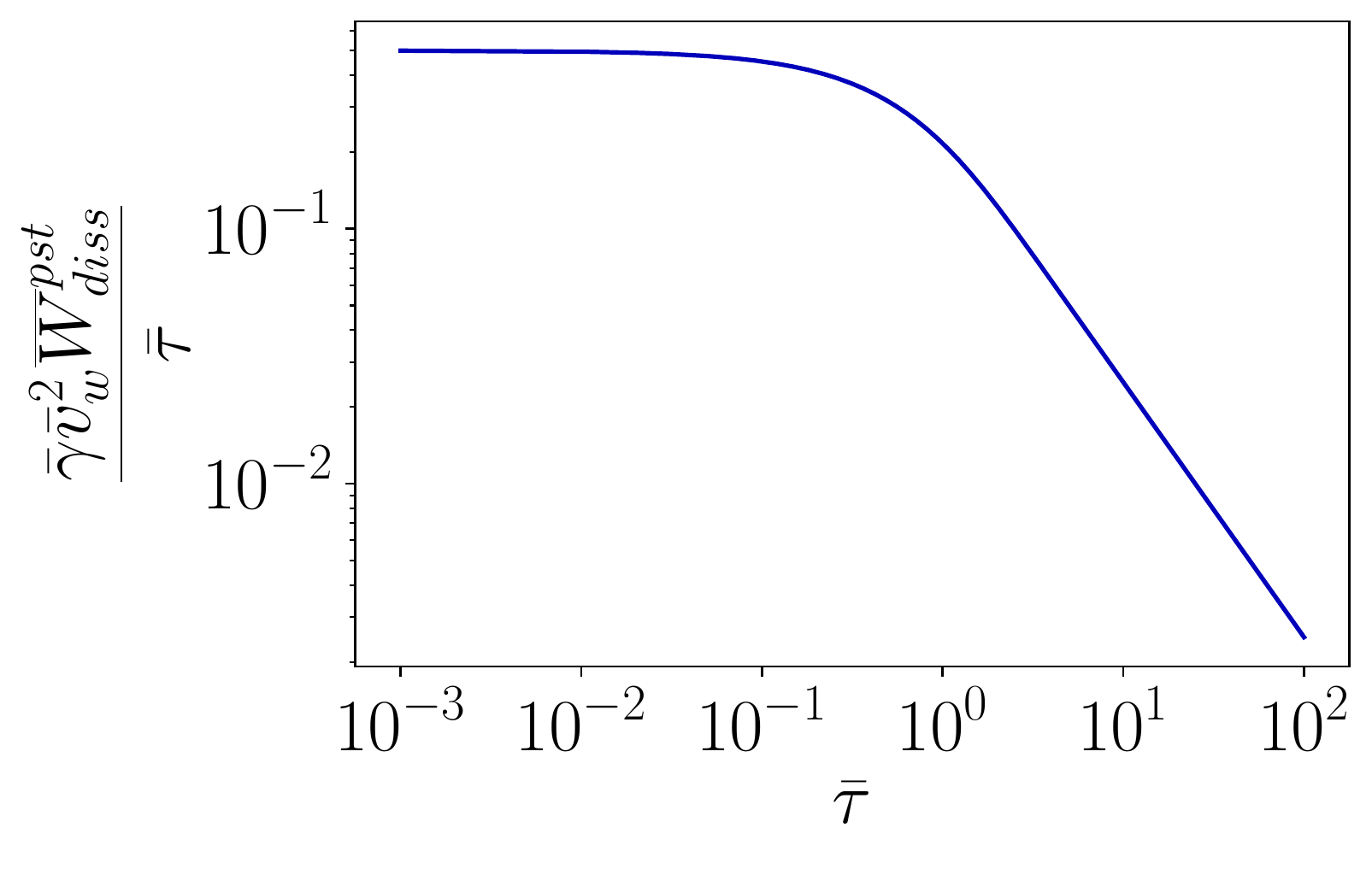}
\caption{Left: magnitude of $\frac{\gamma_w}{\gamma_p} \bar{v}_w^2\overline{W}^{pst}_{\!diss}$ as function of $\frac{\gamma_w}{\gamma_p}$ with $F=0$ and $\bar\tau=0.1,0.5,1,10$ where lighter colors stand for larger values of $\bar\tau$. Right: magnitude of $\frac{\gamma_w}{\gamma_p} \bar{v}_w^2\overline{W}^{pst}_{\!diss}\big/\bar\tau$ as function of $\bar\tau$ for $\gamma_w / \gamma_p=1$ and $\bar{F}=0$ \textit{i.e.} at the maximum value of the left panel. 
\label{fig:app_diss_pst}}
\end{figure}

We comment on the magnitude of $\frac{\gamma_w}{\gamma_p} \bar{v}_w^2\overline{W}^{pst}_{\!diss}$.  For clarity sake, we express it as
\begin{align}
\frac{\gamma_w}{\gamma_p} \bar{v}_w^2\overline{W}^{pst}_{\!diss} = \frac{\gamma_w}{\gamma_p} \frac{1-F/\gamma_p v_0}{1 + \gamma_w/\gamma_p} \left( 1-e^{-2\bar\tau}\right)
\label{eq:app_diss_pst}
\end{align}
Clearly the last expression is maximized at $\bar\tau\rightarrow \infty$ and $F=0$. Accordingly, in the left panel of Fig.\ref{fig:app_diss_pst} we report the values of $\frac{\gamma_w}{\gamma_p} \bar{v}_w^2\overline{W}^{pst}_{\!diss}$ for $F=0$ and $\bar\tau =10$ while in the right panel the values of $\frac{\gamma_w}{\gamma_p} \bar{v}_w^2\overline{W}^{pst}_{\!diss}$ normalized by $\bar\tau$ as function of $\bar\tau$ for $\gamma_w / \gamma_p=1$ and $\bar{F}=0$ \textit{i.e.} at the maximum value of the left panel.

\subsection{Dependence of $\bar\tau_{opt}$ on $\overline{\mathcal{M}}$}\label{app:fig2}

We report the values of $\tau_{opt}$ obtained by the numerical solution of $\partial_{\bar\tau} \langle \tilde\eta\rangle=0$ where $\tilde\eta$ is defined in Eq.~(21) of the main text.
\begin{figure}[h!]
\includegraphics[scale=0.5]{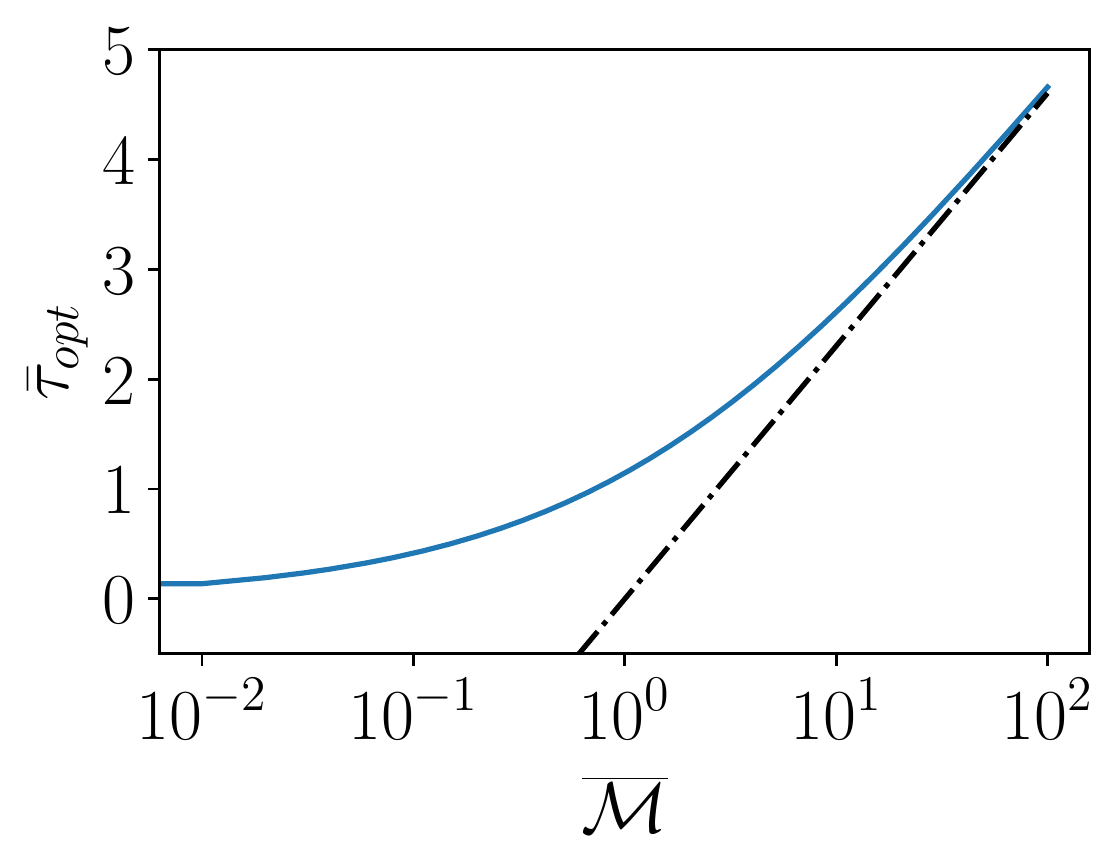}
\caption{Dependence of $\bar\tau_{opt}$ on $\overline{\mathcal{M}}$ for any values of $\bar{F}$ and $\bar\gamma$. The dot-dashed lines stands for $\ln\left(\overline{\mathcal{M}}\right)$.}
\label{fig:app_fig2}
\end{figure}


\end{document}